\documentclass[aps,prd,amsmath,floats,floatfix,twocolumn,
  superscriptaddress,nofootinbib,showpacs]{revtex4-1}
\pdfoutput=1 
\usepackage{amsmath}    
\usepackage{lipsum}
\usepackage{amssymb}
\usepackage{amsfonts}
\usepackage{amsthm}
\usepackage{braket}
\usepackage{bm}
\usepackage{multirow}
\usepackage{color}      
\usepackage{dcolumn}
\usepackage[dvipsnames]{xcolor}
\usepackage{epsfig}
\usepackage{graphicx}   
\usepackage{graphics}
\usepackage[latin1]{inputenc}
\usepackage{latexsym}
\usepackage{mathrsfs}
\usepackage{mathtools}
\usepackage{orcidlink}
\usepackage{rotating}
\usepackage{setspace}
\usepackage{siunitx}
\usepackage{subfigure}  
\usepackage{url}
\usepackage{verbatim}	
\usepackage{hyperref}   
\usepackage{float}
\hypersetup{
    colorlinks=true,
    urlcolor=blue,
    citecolor=blue,
    linkcolor=red
}

\usepackage{xspace} 
\usepackage{cleveref}
\usepackage[toc,page]{appendix}
\usepackage{array}

\usepackage{yfonts}
\usepackage{tikz}
\usetikzlibrary{shapes.geometric, arrows}
\tikzstyle{null} = [rectangle, 
minimum width=0.5cm, 
minimum height=0.5cm, 
text centered, 
fill=white!30]
\tikzstyle{nullblue} = [rectangle, 
minimum width=0.5cm, 
minimum height=0.5cm, 
text centered, 
text = blue,
fill=white!30]
\tikzstyle{nullor} = [rectangle, 
minimum width=0.5cm, 
minimum height=0.5cm, 
text centered, 
text = orange,
fill=white!30]
\tikzstyle{nullred} = [rectangle, 
minimum width=0.5cm, 
minimum height=0.5cm, 
text centered, 
text = red,
fill=white!30]
\tikzstyle{nullsmall} = [rectangle, 
minimum width=0.1cm, 
minimum height=0.1cm]
\tikzstyle{bracket} = [rectangle, 
minimum width=0.3cm, 
minimum height=3cm]
\tikzstyle{ColinStyle} = [rectangle, 
minimum width=1.2cm, 
minimum height=1cm, 
text centered, 
draw=black, 
line width = 0.7pt,
fill=white!30]
\tikzstyle{arrow} = [line width = 1.3 pt,->,>=stealth]
\tikzstyle{line} = [line width = 0.9pt,-,>=stealth]
\tikzstyle{bluethickline} = [line width = 2pt,-,>=stealth, color = blue]
\tikzstyle{orthickline} = [line width = 2pt,-,>=stealth, color = orange]
\tikzstyle{redthickline} = [line width = 2pt,-,>=stealth, color = red]
\makeatletter
\newcommand*{\rom}[1]{\expandafter\@slowromancap\romannumeral #1@}

\newcommand{\Hertz}{{\mbox{\tiny H}}}

\newcommand{\geo}{{\mbox{\tiny geo}}}
\newcommand{\typeD}{{\mbox{\tiny D}}}
\newcommand{\nonD}{{\mbox{\tiny non-D}}}

\newcommand{\even}{{\mbox{\tiny even}}}
\newcommand{\odd}{{\mbox{\tiny odd}}}

\newcommand{\upd}[1]{\textcolor{magenta}{{{#1}}}}

\makeatother
\allowdisplaybreaks

\newcommand{\CAL}{Theoretical Astrophysics 350-17, California Institute of Technology, Pasadena, CA 91125, USA}

\begin{document}
    \title{Spectroscopy of bumpy BHs: non-rotating case}
    \author{Colin Weller\,\orcidlink{0000-0001-5173-5638}}
    \email{cweller@caltech.edu}
    \affiliation{\CAL}
    
    \author{Dongjun Li\,\orcidlink{0000-0002-1962-680X}}
    \email{dlli@caltech.edu}
    \affiliation{\CAL}
    
    \author{Yanbei Chen\,\orcidlink{0000-0002-9730-9463}}
    \affiliation{\CAL}
    
    \date{\today}
    \begin{abstract}
        Recent detections of gravitational waves have made black hole quasinormal modes a powerful tool in testing predictions of general relativity. Understanding the spectrum of these quasinormal modes in a broad class of theories beyond general relativity and a variety of astrophysical environments around black holes remains vital. In this work, we study the quasinormal mode spectrum of parametrized deformations of a non-rotating black hole in the vacuum. Following Vigeland and Hughes, we model these parametrized deformations as axisymmetric multipole moments in the Weyl coordinates with amplitudes much less than the amplitude of the Schwarzschild potential. These tiny bumps in the black hole geometry satisfy the linearized vacuum Einstein equations and are asymptotically flat. We use the recently developed modified Teukolsky formalism to derive one decoupled differential equation for the radiative Weyl scalar $\Psi_0$. We then use the eigenvalue perturbation method to compute the quasinormal mode frequency shifts of both even- and odd-parity modes with $\ell=2,3$ and up to the overtone number $n=2$ for the Weyl multipoles with $\ell_W=2,3$. Our calculation provides an avenue to directly connect the multipole moments of a modified black hole spacetime to the QNM frequency shifts in a parametric way.
    \end{abstract}
    \maketitle
\section{Introduction}\label{sec:introduction}

Recent observations from ground-based laser interferometers \cite{LIGOScientific:2016aoc, LIGOScientific:2018dkp}, pulsar timing arrays \cite{NANOGrav:2023gor}, and very-long-baseline interferometry \cite{EHT:2023fox} have spawned a new era of testing general relativity (GR) \cite{Silva:2022srr, LIGOScientific:2020tif}. While GR has agreed with numerous tests, modifications may emerge beyond some scale \cite{Hawking:1966vg, Penrose:1964wq, Hawking:1976ra, Senovilla:2014gza}. In particular, both GR and the standard model of particle physics fail to explain the existence of dark matter \cite{Petraki:2013wwa}, the accelerating expansion of the universe \cite{Riess:1998cb, Perlmutter:1998np, Amendola:2015ksp}, the present matter-antimatter asymmetry \cite{Canetti:2012zc}, motivating bottom-up modifications to GR to provide explanation. At a more fundamental level, GR and quantum mechanics do not reconcile beyond the Planck scale, which has prompted many to search for a unified theory from the top-down, such as loop quantum gravity \cite{Ashtekar:2021kfp, Rovelli:1997yv}, string theory \cite{Mukhi:2011zz}, and other quantum structure programs \cite{Perez:2012wv}. Both the bottom-up and top-down efforts have resulted in a plethora of beyond-GR (bGR) theories, such as scalar-tensor theories \cite{Sotiriou:2014yhm}, Einstein dilaton Gauss-Bonnet gravity \cite{Gross:1986mw, Kanti:1995vq, Moura:2006pz}, Horndeski theory \cite{Kobayashi:2019hrl}, dynamical Chern-Simons gravity \cite{Jackiw:2003pm, Alexander:2009tp}, $f(R)$ gravity \cite{Sotiriou:2006hs, Sotiriou:2008rp}, and higher-derivative gravity without extra fields \cite{Burgess:2003jk, Donoghue:2012zc, Cano:2019ore}. Given the breadth of bGR theories, it remains critical to test their predictions, especially when gravity is strong and highly dynamic.

One avenue that may constrain bGR theories is examining the gravitational wave (GW) signal of a perturbed, remnant black hole (BH) produced by a binary BH merger. The GWs emitted in the ringdown phase are characterized by the so-called quasinormal modes (QNMs) with complex frequencies. The real part of the QNM frequencies is related to the orbital and precessional frequencies of null geodesics near the light ring, while the imaginary part encodes the Lyapunov exponent of the orbit \cite{Ferrari:1984zz, Cardoso:2008bp, Yang:2012he}. Each QNM can be labeled by three integers, $(\ell,m,n)$, where $\ell$ and $m$ are angular momentum quantum numbers and $n$ denotes the overtone \cite{Leaver:1985ax, Maggiore:2018sht}. In GR, by detecting these QNMs, one can infer the mass and spin of the remnant BH. In bGR theories, these QNMs can carry even more information, such as the length scale of bGR physics and the possible existence of other non-metric fields. Examining the correspondence between QNM spectra and fundamental physics around BHs is generally known as BH spectroscopy \cite{Dreyer:2003bv, Berti:2005ys, Berti:2009kk, Berti:2018vdi}.

Using the linear perturbation theory of a single BH, one can precisely calculate its QNMs in GR. For Schwarzschild BHs obeying spherical symmetry, one can directly separate metric perturbations into even- and odd-parity components, which are well described by the Regge-Wheeler and Zerilli-Moncreif equations \cite{Regge:1957td, Zerilli:1971wd, Moncrief:1974am}. These two equations are one-dimensional Schrodinger-type wave equations in the radial coordinate and are decoupled from each other. The resulting QNMs can be found by imposing the gravitational perturbations to be purely outgoing at infinity and purely ingoing at the horizon \cite{Vishveshwara:1970cc}. For spinning BHs with only axisymmetry, one cannot easily decouple perturbations of different metric components, even in linear order, which makes reducing the Einstein equations into purely radial equations much more challenging. One alternative approach built on the Newman-Penrose (NP) formalism \cite{Newman:1961qr} was developed by Teukolsky and Press \cite{Teukolsky:1973ha, Press:1973zz, Teukolsky:1974yv}, which instead focuses on curvature perturbations represented by perturbations of the Weyl scalars $\Psi_0$ and $\Psi_4$. In this case, one can also find two decoupled ordinary differential equations of $\Psi_0$ and $\Psi_4$ in the radial coordinate. Since then, QNMs in GR have been studied widely in the literature (see Refs.~\cite{Berti:2009kk, Konoplya:2011qq, Nollert:1999ji} for reviews). Many semi-analytic \cite{Leaver:1985ax, Ferrari:1984zz, Fiziev:2011mm, Konoplya:2019hlu} and numerical methods \cite{Gonzalez:2017gwa, Press:1992zz} have been developed to compute QNMs accurately within GR.

Importantly, QNMs in GR are governed by the ``no-hair'' theorem, which requires that a BH is completely determined by its mass, angular momentum, and electric charge \cite{Israel:1967wq, Carter:1971zc, Price:1971fb, Robinson:1975bv}. While it was shown a \textit{real} scalar field cannot source scalar hair \cite{Kleihaus:2023zzs}, it is possible to have ``hairy" solutions when GR is coupled to other fields and in modified gravity (e.g., see \cite{Kleihaus:1997ic, Chrusciel:2012jk}), which has been tested using GW detections and QNMs \cite{Isi:2019aib}. 

An equivalent statement of the no-hair theorem can be given in terms of field multipole moments. Typically, multipole moments describe the expansion of a field that satisfies a linear differential equation, such as Laplace's equation. Nonetheless, Thorne found how to use multipole moments to describe solutions of the Einstein equations by making a post-Newtonian expansion \cite{Thorne:1980ru}. Later, Geroch and Hansen (GH) extended this description to the strong gravity regime \cite{Geroch:1970cd, Geroch:1970nt, Hansen:1974zz}, which is equivalent to Thorne's description when the correct approximation is taken \cite{Yekta:1983}. Similar to the multipoles discovered by Thorne, GH multipoles have two types: mass moments $M_l$ and current moments $S_l$. In particular, for the Kerr geometry, they satisfy (in geometrized units),
\begin{equation}\label{eq:NoHair}
    M_l+i S_l=M(ia)^l\,,\quad a=J/M\,,
\end{equation}
where $J$ and $M$ are the angular momentum and mass of a BH, respectively. These moments were only well-defined for vacuum spacetimes originally, but some recent work has extended the original definition to include sources \cite{Mayerson:2022ekj} and bGR theories \cite{Pappas:2014gca, Cano:2022wwo}. Additionally, it has been shown that given all the GH multipoles for a spacetime, the full metric can be reconstructed \cite{Beig:1980kk, Tahura:2023qqt, Fodor:2020fnq}.
 
The intricate connection between a BH's geometry and its multipole moments led Thorne to advocate GW signals as a way to examine it \cite{Thorne:1992ikj}. Later, Ryan calculated the GWs emitted by an extreme mass-ratio inspiral (EMRI) around an axisymmetric central BH with arbitrary GH multipole moments \cite{Ryan:1995wh}. Following these previous endeavors, Refs.~\cite{Collins:2004ex, Vigeland:2009pr, Vigeland:2011ji} further developed the program of \textit{bumpy} BHs, which deviate from the Kerr geometry by multipole moments (i.e., the ``bumps"), and computed the modifications to EMRI waveforms. However, none of these analyses calculated the shifts in QNM frequencies due to these additional multipole moments. 

On the other hand, there has been a rising interest in performing parametrized ringdown tests \cite{LISA:2022kgy}. Most of the previous studies either focus on non-rotating BHs \cite{Cardoso:2019mqo, Franchini:2022axs} or make the eikonal approximation for rotating BHs \cite{Berti:2018vdi, Glampedakis:2017cgd, Glampedakis:2017dvb, Glampedakis:2019dqh}. However, recent GW detections indicate that most remnant BHs are fast rotating \cite{LIGOScientific:2021djp}, while the eikonal approximation only works properly for QNMs with large $\ell$, which are subdominant in the ringdown signal and may not even work for certain bGR theories \cite{Berti:2018vdi}. Some other efforts consider modifying the potential in the radial Teukolsky equations directly \cite{Volkel:2022aca}, while the map from the QNM frequency shifts to the geometric deformations of a BH becomes not transparent. 

All these challenges in studying the multipole moments of a BH spacetime and performing parametrized ringdown tests motivate this work. In this work, we will apply the modified Teukolsky formalism developed in \cite{Li:2022pcy, Hussain:2022ins, Cano:2023tmv} to directly compute the QNM frequency shifts generated by the multipole moments of the bumpy BHs considered in \cite{Vigeland:2009pr}. This new approach of doing parametrized ringdown tests works in general for Kerr BHs with perturbative axisymmetric deviations. It also does not rely on any eikonal approximation. Furthermore, it allows one to map the deformations of the BH geometry, described by multipole moments, to the QNM frequency shifts directly. Previous works \cite{Cano:2023tmv, Cano:2023jbk, Wagle:2023fwl} have applied the modified Teukolsky formalism to compute the QNMs in specific bGR theories, while we apply it to this theory-agnostic ringdown study. To demonstrate our approach, we will focus on the simplest case of non-rotating bumpy BHs in this work. Although our BHs are non-rotating, the bumps added to them are axisymmetric \cite{Vigeland:2009pr}, so the procedures we implement here work for the rotating case in principle.

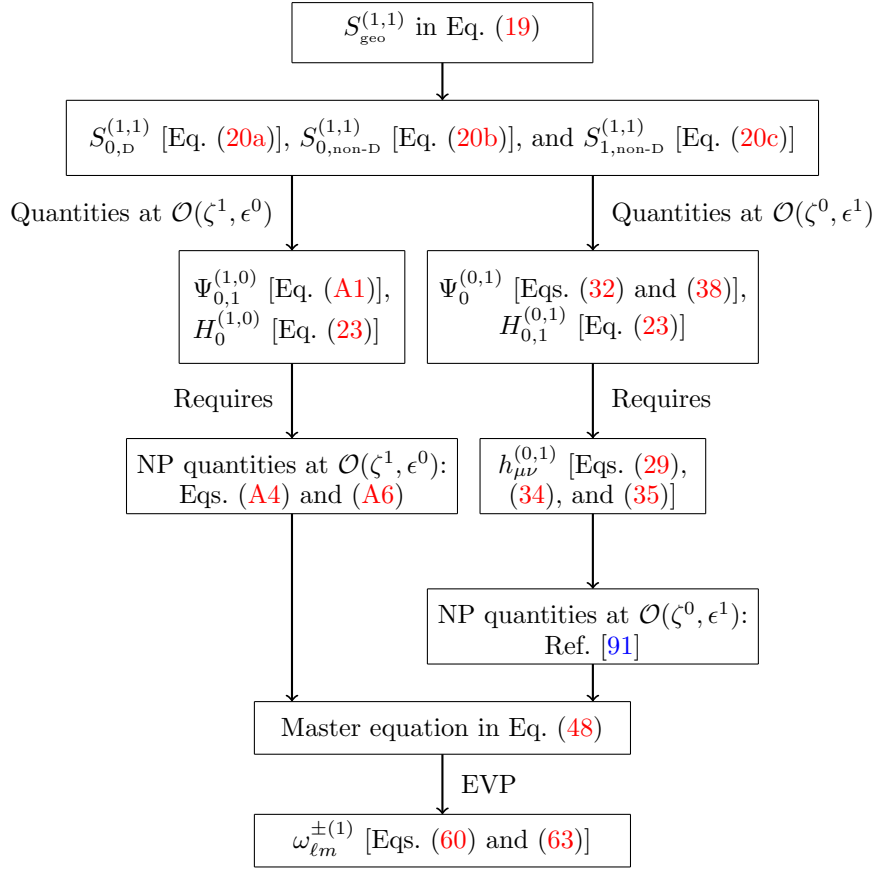
\begin{figure*}
    \centering
    \begin{tikzpicture}
    \node (a) at (0,0) [label=\textrm{$S_{\geo}^{(1,1)}$ in Eq.~\eqref{eq:S_geo}}]{};
    \draw (-2,0) -- (2,0);
    \draw (-2,0.8) -- (2,0.8);
    \draw (-2,0) -- (-2,0.8);
    \draw (2,0) -- (2,0.8);
    \draw [->,thick] (0,0) -- (0,-0.5);
    \node (b) at (0,-1.5) 
    [label=\textrm{$S_{0,\typeD}^{(1,1)}$ [Eq.~\eqref{eq:Sd}],
    $S_{0,\nonD}^{(1,1)}$ [Eq.~\eqref{eq:nond1}],
    and $S_{1,\nonD}^{(1,1)}$ [Eq.~\eqref{eq:nond2}]}]{};
    \draw (-5,-1.5) -- (5,-1.5);
    \draw (-5,-0.5) -- (5,-0.5);
    \draw (-5,-1.5) -- (-5,-0.5);
    \draw (5,-1.5) -- (5,-0.5);
    \draw [->,thick] (-2,-1.5) -- (-2,-2.5);
    \node (ba) at (-2,-2) [label=left:\textrm{Quantities at $\mathcal{O}(\zeta^1,\epsilon^0)$}]{};
    \draw [->,thick] (2,-1.5) -- (2,-2.5);
    \node (bb) at (2,-2) [label=right: \textrm{Quantities at $\mathcal{O}(\zeta^0,\epsilon^1)$}]{};
    
    \node (c1) at (-2,-4) [label={[align=left]
    \textrm{$\Psi_{0,1}^{(1,0)}$ [Eq.~\eqref{eq:psi0_bumpy}],} \\
    \textrm{$H_{0}^{(1,0)}$ [Eq.~\eqref{eq:H_in_Teuk}]}}]{};
    \draw (-3.5,-4) -- (-0.5,-4);
    \draw (-3.5,-2.5) -- (-0.5,-2.5);
    \draw (-3.5,-4) -- (-3.5,-2.5);
    \draw (-0.5,-4) -- (-0.5,-2.5);
    \draw [->,thick] (-2,-4) -- (-2,-5);
    \node (c1a) at (-2,-4.5) [label=left:\textrm{Requires}]{};
    \node (c2) at (-2,-6.2) [label={[align=center]
    \textrm{NP quantities at $\mathcal{O}(\zeta^1,\epsilon^0)$:} \\
    \textrm{Eqs.~\eqref{eq:psi2_bumpy} and \eqref{eq:spin_coeffs_bumpy}}}]{};
    \draw (-4.2,-6) -- (0.2,-6);
    \draw (-4.2,-5) -- (0.2,-5);
    \draw (-4.2,-6) -- (-4.2,-5);
    \draw (0.2,-6) -- (0.2,-5);
    \draw [->,thick] (-2,-6) -- (-2,-8.5);
    
    \node (d1) at (2,-4) [label={[align=center]
    \textrm{$\Psi_{0}^{(0,1)}$ [Eqs.~\eqref{eq:Weylmode}
    and \eqref{eq:Teukolsky_eqns_Schw}],} \\
    \textrm{$H_{0,1}^{(0,1)}$ [Eq.~\eqref{eq:H_in_Teuk}]}}]{};
    \draw (-0.2,-4) -- (4.2,-4);
    \draw (-0.2,-2.5) -- (4.2,-2.5);
    \draw (-0.2,-4) -- (-0.2,-2.5);
    \draw (4.2,-4) -- (4.2,-2.5);
    \draw [->,thick] (2,-4) -- (2,-5);
    \node (d1a) at (2,-4.5) [label=right:\textrm{Requires}]{};
    \node (d2) at (2,-6.2) [label={[align=center]
    \textrm{$h_{\mu\nu}^{(0,1)}$ [Eqs.~\eqref{eq:metric_reconstruct_Schw},} \\
    \textrm{\eqref{eq:Hertz_radial}, and \eqref{eq:reducedop}]}}]{};
    \draw (0.5,-6) -- (3.5,-6);
    \draw (0.5,-5) -- (3.5,-5);
    \draw (0.5,-6) -- (0.5,-5);
    \draw (3.5,-6) -- (3.5,-5);
    \draw [->,thick] (2,-6) -- (2,-7);
    \node (d3) at (2,-8.2) [label={[align=center]
    \textrm{NP quantities at $\mathcal{O}(\zeta^0,\epsilon^1)$:} \\
    \textrm{Ref.~\cite{Wagle:2023fwl}}}]{};
    \draw (-0.2,-8) -- (4.2,-8);
    \draw (-0.2,-7) -- (4.2,-7);
    \draw (-0.2,-8) -- (-0.2,-7);
    \draw (4.2,-8) -- (4.2,-7);
    \draw [->,thick] (2,-8) -- (2,-8.5);
 
    \node (e) at (0,-9.3) [label={[align=center] 
    Master equation in Eq.~\eqref{eq:master_eqn_scheme}}]{};
    \draw (-2.5,-8.5) -- (2.5,-8.5);
    \draw (-2.5,-9.2) -- (2.5,-9.2);
    \draw (-2.5,-8.5) -- (-2.5,-9.2);
    \draw (2.5,-8.5) -- (2.5,-9.2);
    \draw [->,thick] (0,-9.2) -- (0,-10);
    \node (ea) at (0,-9.6) [label=right:\textrm{EVP}]{};
    \node (f) at (0,-10.8) [label={[align=center] 
    \textrm{$\omega_{\ell m}^{\pm(1)}$ 
    [Eqs.~\eqref{eq:inner_product_def} and \eqref{eq:QNM_freq}]}}]{};
    \draw (-2.5,-10) -- (2.5,-10);
    \draw (-2.5,-10.7) -- (2.5,-10.7);
    \draw (-2.5,-10) -- (-2.5,-10.7);
    \draw (2.5,-10) -- (2.5,-10.7);
    \end{tikzpicture}
    \caption{\label{fig:flow_chart}A flow chart describing the procedures and the main equations used for computing the QNM frequency shifts $\omega_{\ell m}^{\pm(1)}$ for bumpy BHs.}
\end{figure*}

The manuscript is organized as follows. In Sec.~\ref{sec:bumpy_BH}, we introduce the specific model of bumpy BHs developed in \cite{Vigeland:2009pr}, explain their motivation for testing GR, and detail their derivation. In Sec.~\ref{sec:MTF}, we review the modified Teukolsky formalism and implement the Chrzanowski-Cohen-Kegeles (CCK) metric reconstruction procedure \cite{Chrzanowski:1975wv, Cohen_Kegeles_1975, Kegeles:1979an, Wald:1978vm, Ori:2002uv} to compute all the necessary source terms in the equation. In Sec.~\ref{sec:EVP}, we review the eigenvalue perturbation (EVP) method in \cite{Zimmerman:2014aha, Mark:2014aja, Hussain:2022ins, Li:2023ulk} and apply it to compute the QNM frequency shifts from the modified Teukolsky equation. We then present the results in Sec.~\ref{sec:results} and discuss future avenues in Sec.~\ref{sec:conclusion}. For convenience, we provide a flow chart in Fig.~\ref{fig:flow_chart} to summarize our procedures for computing the QNM frequency shifts of the bumpy BHs.

\section{Bumpy BHs}
\label{sec:bumpy_BH}
Bumpy BHs were first introduced in \cite{Collins:2004ex} as a way to conveniently parametrize multipole deviations away from the Kerr vacuum. In GR, the no-hair theorem requires a Kerr solution to satisfy 
\begin{equation} \label{eq:GHMoments}
    M_l+iS_l=M(ia)^l\,,
\end{equation}
where $(M_l,S_l)$ are the mass and current multipoles of the BH, respectively. Thus, one can form a null experiment of GR by considering deviations of the form
\begin{equation} \label{eq:DeviationMoments}
    M_l+iS_l=M(ia)^l+\delta M_l+\delta S_l\,.
\end{equation}
In \cite{Vigeland:2009pr}, Vingeland and Hughes further extended the results in \cite{Collins:2004ex} to rectify the nonsmooth nature of the bumps and include spinning bumpy BHs. More specifically, they considerred the Weyl metric \cite{Weyl:1917rtf} 
\begin{equation} \label{eq:WeylMetric}
    ds^2=-e^{2\psi}dt^2+e^{2\gamma-2\psi}
    \left(d\rho^2+dz^2\right)+e^{-2\psi}\rho^2 d\phi^2\,,
\end{equation}
where $(\rho,z)$ are related to the Boyer-Lindquist coordinates $(r,\theta)$ by
\begin{align} \label{eq:changecoord}
    \rho=r\sin\theta \sqrt{1-\frac{2 M}{r}}\,,\quad
    z=(r-M)\cos\theta\,.
\end{align}
The terms $\psi$ and $\gamma$ are functions of $\rho$ and $z$, i.e., $\psi=\psi(\rho,z)$ and $\gamma=\gamma(\rho,z)$. Imposing the spacetime to be Ricci flat, one gets three equations for $\gamma$ and $\psi$:
\begin{subequations} \label{eq:EquMotion}
\begin{align}
    & 0=\frac{\partial^2\psi}{\partial\rho^2}+\frac{1}{\rho}
    \frac{\partial\psi}{\partial\rho}
    +\frac{\partial^2\psi}{\partial z^2}\,, 
    \label{eq:EquMotion_psi} \\
    & \frac{\partial\gamma}{\partial\rho}
    =\rho\left[\left(\frac{\partial \psi}{\partial \rho}\right)^2-\left(\frac{\partial \psi}{\partial z}\right)^2\right]\,, \\
    & \frac{\partial\gamma}{\partial z}
    =2\rho\frac{\partial\psi}{\partial\rho}
    \frac{\partial\psi}{\partial z}\,.
\end{align}    
\end{subequations}

Since Eq.~\eqref{eq:EquMotion_psi} is simply Laplace's equation, $\psi$ can be conveniently chosen as harmonic functions \cite{Vigeland:2009pr}. Reference \cite{Vigeland:2009pr} further perturbatively solved Eq.~\eqref{eq:EquMotion} by expanding the bumpy BH spacetime around the BHs in GR, e.g., 

\begin{align}
    & \psi=\psi_0+\psi_1\,,\; && \psi_1/\psi_0\ll 1\,, \nonumber\\
    & \gamma=\gamma_0+\gamma_1\,,\; && \gamma_1/\gamma_0\ll 1\,.
\end{align}
For non-rotating BHs, $\psi_0$ and $\gamma_0$ correspond to the Schwarzschild metric in the Weyl coordinates with
\begin{equation}
    \begin{aligned}
    \psi_0 &=\ln\tanh(u/2)\,, \\
    \gamma_0 & =-\frac{1}{2}\ln\left(1+\frac{\sin^2 v}{\sinh^2 u}\right)\,.
    \end{aligned}
\end{equation}
where $(u,v)$ are prolate spheroidal coordinates $(u,v)$ and are related to the Weyl coordinates $(\rho,z)$ by 
\begin{subequations}
    \begin{align}
& \rho=M \sinh u \sin v, \\
& z=M \cosh u \cos v.
    \end{align}
\end{subequations}
Since $\psi_1$ is a harmonic function when the spacetime is Ricci flat, it can be decomposed into multipoles in Weyl coordinates, i.e.,
\begin{align} \label{eq:psi_decompostion}
    \psi_{1}(\rho,z)
    =\sum_{\ell_W} B_{\ell_W}M^{\ell_W+1}\frac{Y_{\ell_W 0}(\theta_{\text{Weyl}})}
    {(\rho^2+z^2)^{\frac{\ell_W+1}{2}}}\,,
\end{align}
where $\cos(\theta_{\text{Weyl}})=z/\sqrt{\rho^2+z^2}$, and $B_{\ell_W}$ is a dimensionless coupling constant that parametrizes the magnitude of the bump with multipole $\ell_W$. We can assume that $B_{\ell_W}\ll 1$ for our purposes. Using the coordinate transformation in Eq.~\eqref{eq:changecoord}, one can transform Eq.~\eqref{eq:WeylMetric} in the Weyl coordinates to the Boyer-Lindquist coordinates such that
\begin{align} \label{eq:BumpySchwarz}
    d s^2 =& \;-e^{2 \psi_1}\left(1-\frac{2 M}{r}\right)dt^2
    +e^{2\gamma_1-2\psi_1}\left(1-\frac{2 M}{r}\right)^{-1}dr^2 \nonumber\\
    & \;+r^2 e^{2\gamma_1-2\psi_1}d\theta^2+r^2\sin^2\theta e^{-2\psi_1}d\phi^2.
\end{align}

To solve for the function $\gamma_1$, we need to use the remaining equations of motion in Eq.~\eqref{eq:EquMotion}. Let us focus on $\ell_W=2$ as an example, and the procedures for higher $\ell_W$ are similar. At $\ell_W=2$, Eq.~\eqref{eq:psi_decompostion} gives
\begin{align} \label{eq:psi_l2}
    \psi_1^{\ell_W=2}(\rho, z) 
    =\frac{B_2 M^3}{4} \sqrt{\frac{5}{\pi}} 
    \frac{3\cos^2\theta_{\mathrm{Weyl}}-1}{\left(\rho^2+z^2\right)^{3/2}}\,,
\end{align} 
which in the Boyer-Lindquist coordinates becomes
\begin{align}
    \psi_1^{\ell_W=2}(r,\theta)
    =\frac{B_2 M^3}{4}\sqrt{\frac{5}{\pi}}\frac{1}{d(r,\theta)^3}
    \left[\frac{3(r-M)^2\cos^2\theta}{d(r,\theta)^2}-1\right]\,,
\end{align}
where 
\begin{align} \label{eq:DistanceFunc}
    d(r,\theta) \equiv\left(r^2-2 M r+M^2\cos^2\theta\right)^{1/2}\,.
\end{align}
By using Eq.~\eqref{eq:EquMotion} and imposing that $\gamma_1\to 0$ as $r\to\infty$, one obtains
\begin{align} \label{eq:gamma_l2}
    & \gamma_1^{\ell_W=2}(r,\theta)
    =B_2 \sqrt{\frac{5}{\pi}}\left[\frac{r-M}{2}
    \frac{c_{20}(r)+c_{22}(r)\cos^2\theta}
    {d(r,\theta)^5}-1\right]\,, \nonumber\\
    & c_{20}(r)=2(r-M)^4-5M^2(r-M)^2+3M^4\,, \nonumber\\
    & c_{22}(r)=5M^2(r-M)^2-3M^4\,.
\end{align}
Similarly, at $\ell_W=3$, one can find \cite{Vigeland:2009pr}
\begin{widetext}
\begin{subequations} \label{eq:bumps_l3}
\begin{align}
    & \psi_1^{\ell_W=3}(r,\theta)
    =\frac{B_3 M^4}{4}\sqrt{\frac{7}{\pi}}\frac{1}{d(r,\theta)^4}
    \left[\frac{5(r-M)^3\cos^3\theta}{d(r,\theta)^3}
    -\frac{3(r-M)\cos\theta}{d(r,\theta)}\right]\,, \\
    & \gamma_1^{\ell_W=3}(r,\theta)
    =\frac{B_3 M^5}{2}\sqrt{\frac{7}{\pi}}\cos\theta 
    \left[\frac{c_{30}(r)+c_{32}(r)\cos^2\theta+c_{34}(r)\cos^4\theta
    +c_{36}(r)\cos^6\theta}{d(r,\theta)^7}\right]\,, \\
    & c_{30}(r)=-3r(r-2M)\,,\quad c_{32}(r)=10r(r-2 M)+2 M^2\,,\quad
    c_{34}(r)=-7r(r-2 M)\,,\quad c_{36}(r)=-2M^2\,. \nonumber
\end{align}    
\end{subequations}
\end{widetext}
In this work, we will only focus on the bumps with $\ell_W=2,3$ and compute the corresponding QNM frequency shifts. As one can directly check using Eqs.~\eqref{eq:psi_l2}, \eqref{eq:gamma_l2}, and \eqref{eq:bumps_l3}, the bumps at $\ell_W=2$ and $\ell_W=3$ are of even- and odd-parity, respectively. As we will see later, bumps with different parity will generate QNM spectra with different characteristic structures. Thus, studying the bumps at $\ell_W=2,3$ already allows us to investigate those characteristic features of QNM spectra in the bumpy BH spacetime, while the procedures for computing the QNM spectra of bumps with higher $\ell_W$ are similar.

\section{BH Perturbations in the Modified Teukolsky Formalism}
\label{sec:MTF}

In this section, we briefly review the modified Teukolsky formalism in \cite{Li:2022pcy} and detail how to implement it for the bumpy BHs considered in this work. At the end of this section, we provide the form of the master equations governing the QNMs of these bumpy BHs with the complete expressions provided in the supplementary notebook \cite{ColinBumpyBH}.

\subsection{The modified Teukolsky equation}
\label{sec:modified_Teuk_eqn}

To compute the QNM frequency shifts driven by the bumps in \cite{Vigeland:2009pr}, we choose to apply the modified Teukolsky formalism in \cite{Li:2022pcy, Li:2023ulk, Wagle:2023fwl}. Built upon the seminal work by Teukolsky \cite{Teukolsky:1973ha}, Ref.~\cite{Li:2022pcy} has taken an effective field theory extension of the Teukolsky formalism by using a two-parameter expansion, e.g.,
\begin{equation}
    \Psi_i=\Psi_i^{(0,0)}+\zeta\Psi_i^{(1,0)}
    +\epsilon\Psi_i^{(0,1)}+\zeta\epsilon\Psi_i^{(1,1)}\,,
\end{equation}
where we have taken the Weyl scalars $\Psi_i$ as an example. The coefficient $\zeta$ is a dimensionless parameter that parametrizes the strength of the deviation from GR. In our case, $\zeta=B_{\ell_W}$, where $B_{\ell_W}$ are the amplitudes of the bumps in Eq.~\eqref{eq:psi_decompostion}. The coefficient $\epsilon$ characterizes the magnitude of GW perturbations of certain background spacetime. In this case, the quantities at $\mathcal{O}(\zeta^0,\epsilon^0)$ are evaluated in some GR BH spacetimes of Petrov-type-D. For non-rotating BHs considered in this study, the terms $\Psi_i^{(0,0)}$ are evaluated in the Schwarzschild spacetime. The terms at $\mathcal{O}(\zeta^1,\epsilon^0)$ are driven by bGR modifications to the background spacetime. In this work, these bGR modifications are bumps described by the Weyl multipole potentials $(\psi_1^{\ell_W},\gamma_1^{\ell_W})$ in Sec.~\ref{sec:bumpy_BH}. The terms at $\mathcal{O}(\zeta^0,\epsilon^1)$ correspond to GWs in GR, while the terms at $\mathcal{O}(\zeta^1,\epsilon^1)$ are bGR GWs we are interested in.

Utilizing this expansion, Ref.~\cite{Li:2022pcy} found a set of decoupled differential equations for $\Psi^{(1,1)}_0$ and $\Psi^{(1,1)}_4$, where 
\begin{align} \label{eq:master_eqn_non_typeD_Psi0}
    H_{0}^{(0,0)}\Psi_0^{(1,1)}
    =\mathcal{S}_{\geo}^{(1,1)}+\mathcal{S}^{(1,1)}\,.
\end{align}
In this work, we choose to focus on $\Psi_0^{(1,1)}$, and the equations for $\Psi_4^{(1,1)}$ can be found in \cite{Li:2022pcy}. Here, $H_{0}^{(0,0)}$ is the Teukolsky operator for $\Psi_0$ in GR \cite{Teukolsky:1973ha}. Reference \cite{Li:2022pcy} has divided up the source terms into two categories. The terms in $\mathcal{S}_{\geo}^{(1,1)}$ only depend on the corrections to the background geometry and the GWs in GR, so they are considered purely ``geometrical,'' i.e.,  
\begin{align} 
    \mathcal{S}_{\geo}^{(1,1)}=\mathcal{S}_{0,\typeD}^{(1,1)}
    +\mathcal{S}_{0,\nonD}^{(1,1)}+\mathcal{S}_{1,\nonD}^{(1,1)}\,,  \label{eq:S_geo}
\end{align}
with
\begin{subequations} \label{eq:source_non_typeD_Psi0}
\begin{align}
    \label{eq:Sd} & \mathcal{S}_{0,\typeD}^{(1,1)}=-H_0^{(1,0)}\Psi_0^{(0,1)}\,, \\
    \label{eq:nond1} & \mathcal{S}_{0,\nonD}^{(1,1)}=-H_0^{(0,1)}\Psi_0^{(1,0)}\,, \\
    \label{eq:nond2} & \mathcal{S}_{1,\nonD}^{(1,1)}=H_1^{(0,1)}\Psi_1^{(1,0)}\,.
\end{align}    
\end{subequations}
The other set of source terms is directly driven by the effective stress tensor due to corrections to the Einstein-Hilbert action, i.e.,
\begin{align} \label{eq:S}
    & \begin{aligned} 
        \mathcal{S}^{(1,1)}
        =& \;\mathcal{E}_2^{(0,0)}S_2^{(1,1)}
        +\mathcal{E}_2^{(0,1)}S_2^{(1,0)}
        -\mathcal{E}_1^{(0,0)}S_1^{(1,1)} \\
        & \;-\mathcal{E}_1^{(0,1)}S_1^{(1,0)}\,,
    \end{aligned}
\end{align}
where $S_{1,2}$ are given by 
\begin{subequations} \label{eq:source_bianchi}
\begin{align}
    S_1\equiv 
    & \;\left(\delta-2\bar{\alpha}-2\beta+\bar{\pi}\right)\Phi_{00}
    -\left(D-2\varepsilon-2\bar{\rho}\right)\Phi_{01}\nonumber \\
    & \;+2\sigma\Phi_{10}-2\kappa\Phi_{11}-\bar{\kappa}\Phi_{02}\,,\\
    S_2\equiv 
    & \;\left(\delta-2\beta+2\bar{\pi}\right)\Phi_{01}
    -\left(D-2\varepsilon+2\bar{\varepsilon}-\bar{\rho}\right)\Phi_{02}\nonumber \\
    & \;-\bar{\lambda}\Phi_{00}+2\sigma\Phi_{11}-2\kappa\Phi_{12}\,.
\end{align}    
\end{subequations}
The terms $\Phi_{ab}$ are the NP Ricci scalars, which are projections of the Ricci tensor. The operators $H_{0,1}$, $\mathcal{E}_{0,1}$ are defined as
\begin{equation} \label{eq:H_in_Teuk}
    \begin{aligned}
        & H_0 = \mathcal{E}_2F_2-\mathcal{E}_1 F_1-3\Psi_2\,,\quad
        H_1 = \mathcal{E}_2J_2-\mathcal{E}_1 J_1\,, \\
        & \mathcal{E}_1=E_1-\frac{1}{\Psi_2}\delta\Psi_2 \,,\quad
        \mathcal{E}_2=E_2-\frac{1}{\Psi_2}D\Psi_2\,,
    \end{aligned}
\end{equation}
where $\Psi_2$ is an NP scalar, and we have also defined 
\begin{align} \label{eq:operators_in_Teuk}
    & F_1\equiv\bar{\delta}-4\alpha+\pi\,,\quad
    F_2\equiv\mathbf{\Delta}+\mu-4\gamma\,, \nonumber\\
    & J_1\equiv D-2\varepsilon-4\rho\,, \quad 
    J_2\equiv\delta-2\beta-4\tau\,, \nonumber\\
    & E_1 \equiv\delta-\bar{\alpha}-3\beta+\bar{\pi}-\tau\,, \nonumber\\
    & E_2 \equiv D-3\varepsilon+\bar{\varepsilon}-\rho-\bar{\rho}\,.
\end{align}
Since the non-rotating bumpy BHs considered in this work are Ricci flat \cite{Vigeland:2009pr}, and we do not consider any modifications to the Einstein-Hilbert action, we drop the source term $\mathcal{S}^{(1,1)}$ such that Eq.~\eqref{eq:master_eqn_non_typeD_Psi0} reduces to
\begin{equation}\label{eq:ModTek0S}
    H_0^{(0,0)}\Psi_0^{(1,1)}=\mathcal{S}_{\geo}^{(1,1)}\,.
\end{equation}

To derive Eq.~\eqref{eq:master_eqn_non_typeD_Psi0}, Ref.~\cite{Li:2022pcy} has chosen a gauge following Chandrasekhar \cite{Chandrasekhar_1983} by setting
\begin{align} \label{eq:mtf_gauge}
    \Psi_1^{(0,1)}=\Psi_3^{(0,1)}=\Psi_1^{(1,1)}=\Psi_3^{(1,1)}=0\,.
\end{align}
Similar extensions of the Teukolsky equation have also been developed by \cite{Hussain:2022ins, Cano:2023tmv} without choosing the gauge in Eq.~\eqref{eq:mtf_gauge} and by \cite{Hussain:2022ins} via projecting the Einstein equations following the prescription by Wald in \cite{Wald:1978vm}. For a review of all the NP equations and quantities used in this work as well as the Teukolsky formalism, one can refer to \cite{Newman:1961qr, Teukolsky:1973ha, Press:1973zz, Teukolsky:1974yv, Chandrasekhar_1983, Loutrel:2020wbw, Li:2022pcy}. Before computing the QNMs as solutions of Eq.~\eqref{eq:ModTek0S}, we must first evaluate the operators $H_0^{(0,1)}$ and $H^{(0,1)}_1$, which are driven by the metric of GR GWs.
    
\subsection{Metric reconstruction}
\label{sec:metric_reconstruction}

To solve for the operators $H_{0}^{(0,1)}$ and $H_{1}^{(0,1)}$ at $\mathcal{O}(\zeta^0,\epsilon^1)$, we must compute all the spin coefficients and directional derivatives at this order. All these NP quantities depend on the perturbed metric $h_{\mu\nu}^{(0,1)}$ in GR, so we need to reconstruct $h_{\mu\nu}^{(0,1)}$ from the perturbed Weyl scalar $\Psi_0^{(0,1)}$, which satisfies the GR Teukolsky equation $H^{(0,0)}_0\Psi_0^{(0,1)}=0$. There are multiple approaches for metric reconstruction. For example, one can solve the remaining NP equations systematically after deriving the Teukolsky equation \cite{Chandrasekhar_1983, Loutrel:2020wbw, Ripley:2020xby}. Another more widely used approach is the CCK or CCK-Ori procedure \cite{Chrzanowski:1975wv, Cohen_Kegeles_1975, Kegeles:1979an, Wald:1978vm, Ori:2002uv}. The CCK procedure relies on the radiation gauges. Since $\Psi_0$ characterizes ingoing gravitational radiations of a perturbed BH, it is more convenient to work with the ingoing radiation gauge (IRG), where
\begin{equation}
    h_{ll}^{(0,1)}=h_{lm}^{(0,1)}
    =h_{ln}^{(0,1)}= h^{(0,1)}_{l\Bar{m}}= h_{m\Bar{m}}^{(0,1)}=0\,.
\end{equation}
It was shown in \cite{Price:2006ke} that this gauge can always be imposed on vacuum Petrov-type-D spacetimes. In this case, Refs.~\cite{Chrzanowski:1975wv, Cohen_Kegeles_1975, Kegeles:1979an, Wald:1978vm} found that the metric perturbation solving the linearized Einstein equation in vacuum can be expressed as a smooth second-order differential operator acting on the so-called Hertz potential, $\Psi_{\Hertz}$,
\begin{widetext}
\begin{align} \label{eq:metric_reconstruct}
    h_{\mu\nu}^{(0,1)} 
    =& \;-l_\mu l_\nu(\bar{\delta}+\alpha+3\bar{\beta}-\bar{\tau})
    (\bar{\delta}+4\bar{\beta}+3\bar{\tau})-\bar{m}_\mu\bar{m}_\nu(D-\bar{\rho})(D+3\bar{\rho}) \nonumber\\
    & \;+l_{(\mu}\bar{m}_{\nu)}\left[(D-\bar{\rho}+\rho)(\bar{\delta}+4\bar{\beta}+3\bar{\tau})
    +(\bar{\delta}-\alpha+3\bar{\beta}-\pi-\bar{\tau})
    (D+3\bar{\rho})\right]\bar{\Psi}_{\Hertz} +\textrm{c.c.}\,,
\end{align}
where we have dropped the order-counting superscripts of terms at $\mathcal{O}(\zeta^0,\epsilon^0)$. In the IRG, $\Psi_{\Hertz}$ satisfies the vacuum Teukolsky equation of $\rho^{-4}\Psi_4^{(0,1)}$ \cite{Chrzanowski:1975wv, Cohen_Kegeles_1975, Kegeles:1979an, Wald:1978vm, Ori:2002uv}. On a Schwarzschild background, Eq.~\eqref{eq:metric_reconstruct} simplifies to
\begin{align} \label{eq:metric_reconstruct_Schw}
    h_{\mu\nu}^{(0,1)} 
    =-l_\mu l_\nu(\bar{\delta}+2\beta)(\bar{\delta}+4\beta)
    -\bar{m}_\mu\bar{m}_\nu(D-\rho)(D+3\rho)
    +l_{(\mu}\bar{m}_{\nu)}\left[D(\bar{\delta}+4\beta)
    +(\bar{\delta}+4\beta)(D+3\rho)\right]\bar{\Psi}_{\Hertz}+\textrm{c.c.}\,.
\end{align}
For convenience, let us define
\begin{align} \label{eq:metric_reconstruct_simplify}
    h_{\mu\nu}^{(0,1)}
    =\hat{\mathcal{O}}_{\mu\nu}\bar{\Psi}_\Hertz
    +\hat{\bar{\mathcal{O}}}_{\mu\nu}\Psi_\Hertz\,.
\end{align}
\end{widetext}

To reconstruct NP quantities at $\mathcal{O}(\zeta^0,\epsilon^1)$, we start from the perturbed tetrad at $\mathcal{O}(\zeta^0,\epsilon^1)$. It was found in \cite{Campanelli:1998jv, Loutrel:2020wbw} that in the IRG, one can pick the tetrad at $\mathcal{O}(\zeta^0,\epsilon^1)$ to be
\begin{align} \label{eq:perturbed_tetrad}
    & l^{\mu(0,1)}=0\,,\quad
    n^{\mu(0,1)}
    =\frac{1}{2}h_{nn}^{(0,1)}l^{\mu}\,,\nonumber\\
    & m^{\mu(0,1)}=h_{nm}^{(0,1)}l^{\mu}
    -\frac{1}{2}h_{mm}^{(0,1)}\bar{m}^{\mu}\,.
\end{align}
To compute spin coefficients from Eq.~\eqref{eq:perturbed_tetrad}, we can use the linearized commutation relations. For Weyl scalars, one can directly evaluate the Riemann tensor and project it onto the NP basis. A better approach is to use the Ricci identities \cite{Campanelli:1998jv, Loutrel:2020wbw}, which express Weyl scalars in terms of derivatives of spin coefficients, so we can reuse the reconstructed spin coefficients. More detailed procedures and the corresponding results of the reconstructed NP quantities in GR can be found in \cite{Campanelli:1998jv, Loutrel:2020wbw, Wagle:2023fwl}. We will directly use these results in \cite{Campanelli:1998jv, Loutrel:2020wbw, Wagle:2023fwl} and not present them here for simplicity.

After expressing all the NP quantities in terms of the Hertz potential $\Psi_{\Hertz}$, the next step is to calculate $\Psi_{\Hertz}$ from the perturbed Weyl scalar $\Psi_0^{(0,1)}$. One approach, developed by Ori in \cite{Ori:2002uv}, is using the Teukolsky-Starobinsky identity \cite{Teukolsky:1974yv, Starobinsky:1973aij, Starobinskil:1974nkd} to invert the equation expressing $\Psi_0^{(0,1)}$ in terms of $\bar{\Psi}_{\Hertz}$. Decomposing the $(\ell,m)$ mode of $\Psi_0^{(0,1)}$ and $\bar{\Psi}_{\Hertz}^{(0,1)}$ as
\begin{align} 
    & \Psi^{(0,1)}_{0,\ell m}
    ={}_{2}R_{\ell m}^{(0,1)}(r){}_{2}\mathcal{Y}_{\ell m}(\theta,\phi)
    e^{-i\omega_{\ell m}t}\,, \label{eq:Weylmode} \\
    & \bar{\Psi}_{\Hertz,\ell m}
    ={}_{2}\hat{R}_{\ell m}(r){}_{2}\mathcal{Y}_{\ell m}(\theta,\phi)
    e^{-i\omega_{\ell m}t}\,,
\end{align}
Ref.~\cite{Ori:2002uv} found that
\begin{equation} \label{eq:Hertz_radial}
    {}_{2}\hat{R}_{\ell m}(r)
    =-\frac{2}{\mathfrak{C}}\Delta^2(r)(D^\dagger_{\ell m})^4 \left[\Delta^2(r)\,{}_{2}R_{\ell m}^{(0,1)}(r) \right]\,, 
\end{equation}
where $\Delta(r)\equiv r^2-2Mr+\chi^2M^2$ and
\begin{align} \label{eq:reducedop}
    & D_{\ell m}
    =\partial_r+i\frac{am-(r^2+a^2)\omega_{\ell m}}{\Delta(r)}\,, \nonumber\\
    & D^\dagger_{\ell m}
    =\partial_r-i\frac{am-(r^2+a^2)\omega_{\ell m}}{\Delta(r)}\,.
\end{align}
Here, ${}_{s}\mathcal{Y}_{\ell m}(\theta,\phi)\equiv{}_{s}S_{\ell m}(\theta)e^{im\phi}$ and ${}_{s}R_{\ell m}^{(0,1)}(r)$ are solutions to the angular and radial Teukolsky equations, respectively,
\begin{widetext}
\begin{subequations} \label{eq:Teukolsky_eqns}
\begin{align}
    & \left[\frac{1}{\sin\theta}\frac{d}{d\theta}\left(\sin\theta\frac{d}{d\theta}\right)
    -\gamma_{\ell m}^2\sin^2\theta-\frac{(m+s\cos\theta)^2}{\sin^2\theta}
    -2\gamma_{\ell m} s\cos\theta+s
    +2m\gamma_{\ell m}+{}_s\lambda_{\ell m}\right]{}_sS_{\ell m}(\theta)=0\,, 
    \label{eq:angular_Teukolsky_eqn}\\
    & \left[\Delta(r)^{-s}\frac{d}{dr}\left(\Delta(r)^{s+1}\frac{d}{dr}\right)
    +\frac{K^2(r)-2is(r-M)K(r)}{\Delta(r)}+4is\omega_{\ell m} r
    -{}_{s}\lambda_{\ell m}\right]{}_{s}R_{\ell m}^{(0,1)}(r)=0\,,
    \label{eq:radial_Teukolsky_eqn}
\end{align}   
\end{subequations}
where
\begin{equation} 
    \gamma_{\ell m}\equiv \chi M\omega_{\ell m}\,,\quad
    {}_{s}\lambda_{\ell m}\equiv{}_{s}A_{\ell m}+s+|s|\,,\quad
    K(r)\equiv\left(r^2+\chi^2M^2\right)\omega_{\ell m}-\chi M m\,,
\end{equation}
with ${}_{s}A_{\ell m}$ being the Teukolsky's angular separation  constant~\cite{Teukolsky:1973ha}. For non-rotating BHs we are considering in this paper, Eq.~\eqref{eq:Teukolsky_eqns} reduces to
\begin{subequations} \label{eq:Teukolsky_eqns_Schw}
\begin{align} 
    & \left[\frac{1}{\sin\theta}\frac{d}{d\theta}\left(\sin\theta\frac{d}{d\theta}\right)-\frac{(m+s\cos\theta)^2}{\sin^2\theta}+s
    +(\ell-1)(\ell+s+1)\right]{}_sS_{\ell m}(\theta)=0\,, \\
    & \left[\Delta(r)^{-s}\frac{d}{dr}\left(\Delta(r)^{s+1}\frac{d}{dr}\right)
    +\frac{\omega_{\ell m}r(\omega_{\ell m}r^2-2is(r-M))}{r-2M}
    +4is\omega_{\ell m} r-(\ell-1)(\ell+s+1)\right]{}_{s}R_{\ell m}^{(0,1)}(r)=0\,.
\end{align}   
\end{subequations}
The constant $\mathfrak{C}$ is the Teukolsky-Starobinsky coefficient \cite{Teukolsky:1974yv, Starobinsky:1973aij, Starobinskil:1974nkd, Ori:2002uv, Cano:2023tmv, Cano:2023jbk},
\begin{align}
    \mathfrak{C}
    =& \;144M^2\omega_{\ell m}^2
    +\left(8+6{}_{s}B_{\ell m}+{}_{s}B_{\ell m}^2\right)^2
    -8\left[-8+{}_{s}B_{\ell m}^2
    \left(4+{}_{s}B_{\ell m}\right)\right]m\gamma_{\ell m} \nonumber\\ 
    & \;+4\left[8-2{}_{s}B_{\ell m}-{}_{s}B_{\ell m}^2+{}_{s}B_{\ell m}^3
    +2\left(-2+{}_{s}B_{\ell m}\right)
    \left(4+3{}_{s}B_{\ell m}\right)m^2\right]\gamma_{\ell m}^2 \nonumber\\
    & \;-8m\left(8-12{}_{s}B_{\ell m}+3{}_{s}B_{\ell m}^2
    +4\left(-2+{}_{s}B_{\ell m}\right)m^2\right)
    \gamma_{\ell m}^3 \nonumber \\ 
    & \;+2\left(42-22{}_{s}B_{\ell m}+3{}_{s}B_{\ell m}^2
    +8\left(-11+3{}_{s}B_{\ell m}\right)m^2
    +8m^4\right)\gamma_{\ell m}^4 \nonumber\\ 
    & \;-8 m\left[3{}_{s}B_{\ell m}
    +4\left(-4+m^2\right)\right]\gamma_{\ell m}^5
    +4\left(-7+{}_{s}B_{\ell m}+6m^2\right)\gamma_{\ell m}^6
    -8m\gamma_{\ell m}^7+\gamma_{\ell m}^8\,,
\end{align}
where $\tilde{\alpha}=a^2-am/\omega_{\ell m}$ and ${}_{s}B_{\ell m}={}_{s}{A}_{\ell m}+s$. Notice that we have used the Teukolsky-Starobinsky coefficient found in \cite{Cano:2023tmv, Cano:2023jbk} instead of the original one in \cite{Teukolsky:1974yv, Starobinsky:1973aij, Starobinskil:1974nkd, Ori:2002uv}, the latter of which was derived for real frequencies.
\end{widetext}

Finally, since we have chosen the gauge in Eq.~\eqref{eq:mtf_gauge}, where $\Psi_1^{(0,1)}=\Psi_3^{(0,1)}=0$, we need to perform additional tetrad rotations to remove $\Psi_{1,3}^{(0,1)}$. The required tetrad rotations and the transformation of NP quantities under tetrad rotations are provided in \cite{Wagle:2023fwl}. Using the procedures above, we compute the NP quantities at $\mathcal{O}(\zeta^0,\epsilon^1)$ for Schwarzschild BHs and evaluate the operators $H^{(0,1)}_0$ and $H^{(0,1)}_1$ in the ``geometrical" source term \upd{$\mathcal{S}_{\geo}^{(1,1)}$}. The results can be found in the supplemental material \cite{ColinBumpyBH}, and we present the perturbed metric $h_{ab}^{(0,1)}$ in the NP basis here as an example,
\begin{widetext}
\begin{subequations}
\begin{align}
    h_{nn}^{(0,1)}
    =& \;\frac{\sqrt{\ell(\ell+1)(\ell^2+\ell-2)}}{2r^2}
    {}_{2}\hat{R}_{\ell m}(r){}_{0}Y_{\ell m}(\theta,\phi)
    e^{-i\omega_{\ell m}t}+\textrm{c.c.}\,, \\
    h_{nm}^{(0,1)}
    =& \;\frac{\sqrt{\ell^2+\ell-2}}{\sqrt{2}r^2(r-2M)}
    \left[r(r-2M)\partial_r-r(2+i\omega_{\ell m}r)+4M\right]
    {}_{2}\hat{R}_{\ell m}(r){}_{1}Y_{\ell m}(\theta,\phi)
    e^{-i\omega_{\ell m}t}\,, \\
    h_{mm}^{(0,1)}
    =& \;\frac{1}{r(r-2M)^2}
    \left[2(r-2M)(M-i\omega_{\ell m}r^2)\partial_r
    +(r-2M)(\ell^2+\ell-2+6i\omega_{\ell m}r)
    -2i\omega_{\ell m}r(M-i\omega_{\ell m}r^2)\right] \nonumber\\
    & \;{}_{2}\hat{R}_{\ell m}(r){}_{2}Y_{\ell m}(\theta,\phi)
    e^{-i\omega_{\ell m}t}\,.
\end{align}
\end{subequations}
In the next subsection, we will compute the NP quantities at $\mathcal{O}(\zeta^1,\epsilon^0)$ so we can evaluate the modified Teukolsky equation of $\Psi_0^{(1,1)}$ in Eq.~\eqref{eq:ModTek0S}.
\end{widetext}

\subsection{The modified Teukolsky equation for non-rotating bumpy BHs}
\label{sec:modified_Teuk_eqn_bumpy_BH}

To evaluate Eq.~\eqref{eq:ModTek0S}, we also need to compute the NP quantities at $\mathcal{O}(\zeta^0,\epsilon^0)$ and $\mathcal{O}(\zeta^1,\epsilon^0)$. For NP quantities at $\mathcal{O}(\zeta^0,\epsilon^0)$, we evaluate them on the Schwarzschild background and use the Kinnersely tetrad,
\begin{subequations} \label{eq:BackTetrad}
\begin{align}
    & l^{\mu(0,0)}
    =\left(\frac{r^2}{r(r - 2M) },1,0,0\right)\,, \\
    & n^{\mu(0,0)}
    =\left(\frac{1}{2},-\frac{r(r-2M)}{2r^2},0,0\right)\,, \\
    & m^{\mu(0,0)}
    =\frac{1}{\sqrt{2}r}\left(0,0,1, i \csc{\theta}\right)\,,
\end{align}
\end{subequations}
where all the NP quantities at $\mathcal{O}(\zeta^0,\epsilon^0)$ can be found in \cite{Teukolsky:1973ha, Chandrasekhar_1983}. At $\mathcal{O}(\zeta^1,\epsilon^0)$, let us first linearize Eq.~\eqref{eq:BumpySchwarz} since we assume the dimensionless amplitude $B_{\ell_W}$ of each bump satisfies $B_{\ell_W}\ll 1$. In this case, Eq.~\eqref{eq:BumpySchwarz} becomes
\begin{align} \label{eq:LinearPerturbMetric}
    & h^{(1,0)}_{\mu \nu}dx^{\mu}dx^{\nu} \nonumber\\
    =& \;-2\psi_1\left(1-\frac{2 M}{r}\right)dt^2
    +(2\gamma_1-2\psi_1)\left(1-\frac{2M}{r}\right)^{-1}dr^2 \nonumber\\
    & \;+r^2(2\gamma_1-2\psi_1)d\theta^2-2\psi_1r^2\sin^2\theta d\phi^2\,.
\end{align}
Using Eq.~(51) of \cite{Li:2022pcy}, we find the following tetrad satisfies all the orthogonality conditions,
\begin{subequations} \label{eq:StationaryTet}
\begin{align}
    & l^{\mu(1,0)}
    =\left(-\frac{r\psi_1}{r-2M},\psi_1-\gamma_1, 0, 0 \right)\,,\\
    & n^{\mu(1,0)}
    =\left(-\frac{1}{2}\psi_1,\frac{(r-2M)\left(\gamma_1
    -\psi_1\right)}{2r}, 0, 0\right)\,,\\
    & m^{\mu(1,0)}
    =\left(0,0,\frac{\psi_1-\gamma_1}{\sqrt{2}r},
    \frac{i\csc{\theta}\psi_1}{\sqrt{2}r}\right)\,.
\end{align}
\end{subequations}
Using this tetrad, we can compute Weyl scalars and spin coefficients at $\mathcal{O}(\zeta^1,\epsilon^0)$. These quantities are summarized in Appendix~\ref{appendix:order_10_quantities}. The differential operator $H^{(1,0)}_0$ is provided in the supplemental material \cite{ColinBumpyBH}.

Now, we have all the ingredients to evaluate the modified Teukolsky equation in Eq.~\eqref{eq:ModTek0S}. Due to the complication of the full equation, we choose to provide it in the supplemental material \cite{ColinBumpyBH}, and we present here the schematic form of the equation for the convenience of computing the QNM frequencies. First, since all the terms in $\mathcal{S}_{\geo}^{(1,1)}$ depend on the perturbed metric $h_{\mu\nu}^{(0,1)}$ linearly via $\Psi_0^{(0,1)}$, $H_0^{(0,1)}$, or $H_1^{(0,1)}$, we can write $\mathcal{S}_{\geo}^{(1,1)}$ as

\begin{equation}\label{eq:SigmaIntro}
    \mathcal{S}_{\geo}^{(1,1)}=\hat{\Sigma}^{\mu\nu(1,0)}h^{(0,1)}_{\mu\nu}\,,
\end{equation}
where $\hat{\Sigma}^{\mu\nu(1,0)}$ is determined by the reconstructed NP quantities found in Sec.~\ref{sec:metric_reconstruction} and the modifications of NP quantities due to the bumps. For convenience, we will drop the superscript for order counting of $\hat{\Sigma}^{\mu\nu(1,0)}$. Using Eqs.~\eqref{eq:metric_reconstruct} and \eqref{eq:Hertz_radial}, we can write \eqref{eq:SigmaIntro} as
\begin{align}\label{eq:SourceExpand}
    H^{(0,0)}_0\Psi^{(1,1)}_0 
    =\hat{\Sigma}^{\mu\nu}\left(\hat{\mathcal{O}}_{\mu\nu}+ \hat{\bar{\mathcal{O}}}_{\mu\nu}\hat{\mathcal{C}}\right)\hat{\mathcal{D}}\Psi^{(0,1)}_0\,,
\end{align}
where $\hat{\mathcal{C}}$ is the complex conjugate operator, and $\hat{\mathcal{D}}$ is an operator that satisfies $\hat{\mathcal{D}}\Psi^{(0,1)}_0=\bar{\Psi}_\Hertz$, which can be determined from Eq.~\eqref{eq:Hertz_radial}. Expanding $\Psi^{(0,1)}_0$ and $\Psi^{(1,1)}_0$ as
\begin{align}
    & \Psi_{0}^{(0,1)}
    =\sum_{\ell,m}{}_{2}\psi_{\ell m}^{(0,1)}(r,\theta)
    e^{-i\omega_{\ell m}t+im\phi}\,, \\
    & \Psi_{0}^{(1,1)}
    =\sum_{\ell,m}{}_{2}\psi_{\ell m}^{(1,1)}(r,\theta)
    e^{-i\omega_{\ell m}t+im\phi}\,,
\end{align}
we get the mode decomposition of Eq.~\eqref{eq:SourceExpand} to be
\begin{widetext}
\begin{align} \label{eq:master_eqn_scheme}
    \sum_{\ell,m} H_{0,\ell m}\left[{}_{2}\psi_{\ell m}^{(1,1)}(r,\theta)\right]
    e^{-i\omega_{\ell m} t+im\phi}
    =& \;\sum_{\ell,m}P_{\ell m}\left[{}_{2}\psi_{\ell m}^{(0,1)}(r,\theta)\right]
    e^{-i\omega_{\ell m} t+im\phi}
    +Q_{\ell m}\left[{}_{2}\bar{\psi}_{\ell m}^{(0,1)}(r,\theta)\right]
    e^{i\bar{\omega}_{\ell m}t-im\phi}\,,
\end{align}
\end{widetext}
where $H_{0,\ell m}$ is the $(\ell,m)$ mode of the Teukolsky operator $H_{0}$ for $\Psi_0$ in GR. $P_{\ell m}$ and $Q_{\ell m}$ are operators depending on the coordinates $(r,\theta)$ and acting on the Weyl scalar perturbation $\Psi_0^{(0,1)}$ in GR. In particular,
\begin{align}
    P_{\ell m}=\left(\hat{\Sigma}^{\mu\nu}\hat{\mathcal{O}}_{\mu\nu}
    \hat{\mathcal{D}}\right)_{\ell m}\,,\quad
    Q_{\ell m}=\left(\hat{\Sigma}^{\mu\nu}\hat{\bar{\mathcal{O}}}_{\mu\nu} 
    \hat{\mathcal{D}}\right)_{\ell m}\,.
\end{align}
One may further decompose ${}_{2}\psi_{\ell m}^{(0,1)}(r,\theta)$ and ${}_{2}\psi_{\ell m}^{(1,1)}(r,\theta)$ into spin-weighted spheroidal harmonics
\begin{subequations}
\begin{align}
    & {}_{2}\psi_{\ell m}^{(0,1)}(r,\theta)
    ={}_{2}R_{\ell m}^{(0,1)}(r){}_{2}S_{\ell m}(\theta)\,,\\
    & {}_{2}\psi_{\ell m}^{(1,1)}(r,\theta)
    ={}_{2}R_{\ell m}^{(1,1)}(r){}_{2}S_{\ell m}(\theta)\,.
\end{align}
\end{subequations}
where ${}_{s}R_{\ell m}^{(0,1)}(r)$ and ${}_{s}S_{\ell m}(\theta)$ are radial and angular Teukolsky functions for a spin $s$ particle in GR, respectively, and they satisfy Eq.~\eqref{eq:Teukolsky_eqns}. One thing we notice in Eq.~\eqref{eq:master_eqn_scheme} is that the $(\ell,m)$ and $(\ell,-m)$ modes are coupled to each other in the source terms on the right-hand side, which means we actually need to solve these two modes jointly. In the next section, we will show how to solve these two modes jointly and compute the QNM frequency shifts following the prescription in \cite{Li:2023ulk}.

\section{EVP}
\label{sec:EVP}

In this section, we prescribe the procedures to calculate the QNM frequency shifts due to those bumps using Eq.~\eqref{eq:master_eqn_scheme}. From Eq.~\eqref{eq:master_eqn_scheme}, we notice that the GR QNMs are resonantly driving the modified Teukolsky equation since the homogeneous part of Eq.~\eqref{eq:master_eqn_scheme} is the same as the one in GR, potentially leading to secularly growing terms. One solution is to perform a multiple-scale analysis \cite{bender2013advanced}, which was employed to study spin-precessing systems and post-Newtonian dynamics in GR \cite{Klein:2013qda, Gerosa:2015tea, Chatziioannou:2017tdw}. Another solution is the Poincar\'{e}-Lindstedt method, which leverages shifts of the eigenfrequency to cancel off secularly growing terms. In this case, the shift in the eigenfrequency plays a similar role as the slow timescale in the multiple-scale analysis. Following a similar idea, Refs.~\cite{Zimmerman:2014aha, Mark:2014aja} developed the EVP method by perturbing the QNM frequency in GR, i.e.,
\begin{equation} \label{eq:omega_expansion}
    \omega_{\ell m}=\omega_{\ell m}^{(0)}+\zeta\omega_{\ell m}^{(1)}\,.
\end{equation}
In GR, the QNM frequencies $\omega_{lm}^{(0)}$ satisfy the following symmetry \cite{Leaver:1985ax}
\begin{equation} \label{eq:omega0_symmetry}
    \omega^{(0)}_{\ell m}=-\bar{\omega}^{(0)}_{\ell -m}\,.
\end{equation}
As shown in detail in \cite{Li:2023ulk}, to solve Eq.~\eqref{eq:master_eqn_scheme} consistently, one has to assume that the same symmetry still holds at the bGR level, i.e.,
\begin{equation} \label{eq:omega1_symmetry}
    \omega^{(1)}_{\ell m}=-\bar{\omega}^{(1)}_{\ell -m}\,.
\end{equation}
In this case, all the source terms in Eq.~\eqref{eq:master_eqn_scheme} are either proportional to $e^{-i\omega_{\ell m}t}$ or $e^{-i\omega_{\ell -m}t}$. Since the $(\ell,m)$ and $(\ell,-m)$ modes are coupled to each other, we need to solve these two modes jointly by focusing on the linear combination \cite{Hussain:2022ins, Li:2023ulk, Wagle:2023fwl},
\begin{align} \label{eq:solution_ansatz}
    & \Psi_{0,\ell m}^{\eta(0,1)}
    =\Psi_{0,\ell m}^{(0,1)}
    +\eta_{\ell m}\Psi_{0,\ell -m}^{(0,1)}\,, \nonumber\\
    & \Psi_{0,\ell m}^{\eta(1,1)}
    =\Psi_{0,\ell m}^{(1,1)}
    +\eta_{\ell m}\Psi_{0,\ell -m}^{(1,1)}\,.
\end{align}
The constant $\eta_{\ell m}$ is the relative coefficient between the $(\ell,m)$ and $(\ell,-m)$ modes, which is well-defined if we fix the normalization of ${}_{2}\psi_{\ell m}^{(0,1)}(r,\theta)$ to satisfy
\begin{align} \label{eq:parity_Teuk_func}
    & {}_{s}\bar{R}_{\ell m\omega}^{(0,1)}(r)
    =(-1)^{m}{}_{s}R_{\ell-m-\bar{\omega}}^{(0,1)}(r)\,, \nonumber\\
    & {}_{s}S_{\ell m\omega}(\pi-\theta)
    =(-1)^{m+\ell}{}_{-s}S_{\ell m\omega}(\theta)\,, \nonumber\\
    & {}_{s}\bar{S}_{\ell m\omega}(\theta)
    =(-1)^{m+s}{ }_{-s}S_{\ell-m-\bar{\omega}}(\theta)\,.
\end{align}
\begin{figure}[t!]
\includegraphics[width=0.3\textwidth]{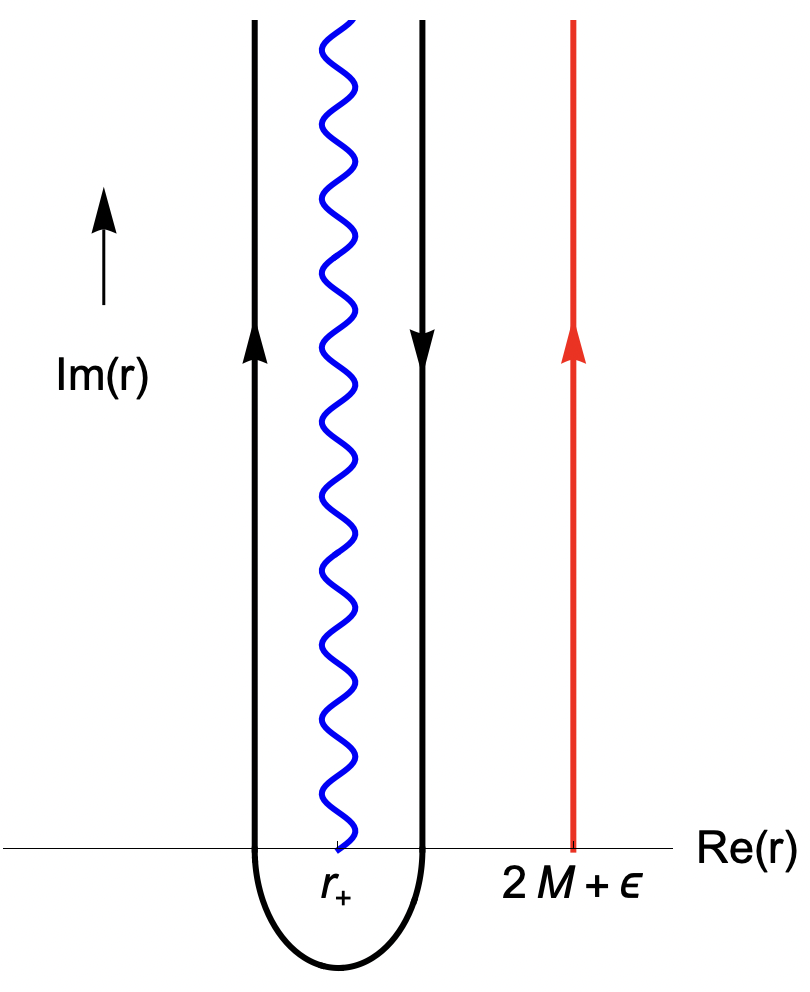}
    \caption{The contour $\mathscr{C}_1$ that wraps around the QNM wavefunction branch cut parallel to the imaginary axis at the horizon $r_{+}=2M$. The modified contour $\mathscr{C}_2$ is also shown, which extends from $r=2M+\epsilon$ to $r=2M+\epsilon+i\infty$.} 
    \label{fig:NewContour}
\end{figure}
After plugging the ansatz in Eq.~\eqref{eq:solution_ansatz} into the Teukolsky equation of $\Psi_0$ in GR [Eq.~\eqref{eq:Teukolsky_eqns_Schw}] and the modified Teukolsky equation [Eq.~\eqref{eq:master_eqn_scheme}], matching the phase of the terms, and expanding the equations over $\zeta$, we get
\begin{widetext}
\begin{subequations} \label{eq:master_eqn_scheme2}
\begin{align}
    H_{0,\ell m}\left[{}_{2}\psi_{\ell m}^{(1,1)}(r,\theta)\right]
    +\omega_{\ell m}^{(1)}\partial_\omega H_{0,\ell m}
    \left[{}_{2}\psi_{\ell m}^{(0,1)}(r,\theta)\right]
    =& \;P_{\ell m}\left[{}_{2}\psi_{\ell m}^{(0,1)}(r,\theta)\right]
    +\bar{\eta}_{\ell m}Q_{\ell-m}
    \left[{}_{2}\bar{\psi}_{\ell -m}^{(0,1)}(r,\theta)\right]\,, 
    \label{eq:master_eqn_scheme2_lm} \\
    \eta_{\ell m}H_{0,\ell-m}
    \left[{}_{2}\psi_{\ell-m}^{(1,1)}(r,\theta)\right]
    +\eta_{\ell m}\omega_{\ell-m}^{(1)}\partial_\omega H_{0,\ell-m}
    \left[{}_{2}\psi_{\ell-m}^{(0,1)}(r,\theta)\right]
    =& \;\eta_{\ell m}P_{\ell-m}
    \left[{}_{2}\psi_{\ell-m}^{(0,1)}(r,\theta)\right]
    +Q_{\ell m}\left[{}_{2}\bar{\psi}_{\ell m}^{(0,1)}(r,\theta)\right]\,, \label{eq:master_eqn_scheme2_l-m}
\end{align}
\end{subequations}
\end{widetext}
where the second term on the left-hand side of Eqs.~\eqref{eq:master_eqn_scheme2_lm} and \eqref{eq:master_eqn_scheme2_l-m} comes from expanding $\omega_{\ell m}$ about $\zeta$ [i.e., Eq.~\eqref{eq:omega_expansion}] in the GR Teukolsky equation $H_{0,\ell m}\left[{}_{2}\psi_{\ell m}^{(0,1)}(r,\theta)\right]=0$. All the $\omega_{\ell m}$ and $\omega_{\ell-m}$ terms in Eq.~\eqref{eq:master_eqn_scheme2} are evaluated on the GR QNM frequencies $\omega_{\ell m}^{(0)}$ and $\omega_{\ell-m}^{(0)}$, respectively. Furthermore, following \cite{Li:2023ulk}, one can apply a parity-complex conjugate transformation $\hat{\mathcal{P}}$, 
\begin{equation} \label{def:cal_P}
    \hat{\mathcal{P}}f(\theta,\phi)\equiv\bar{f}(\pi-\theta,\phi+\pi)\,,
\end{equation}
on Eq.~\eqref{eq:master_eqn_scheme2_l-m} such that it becomes
\begin{widetext}
\begin{equation} \label{eq:master_eqn_scheme3_l-m}
\begin{aligned}
    & (-1)^m\bar{\eta}_{\ell m}H_{0,\ell m}
    \left[{}_{2}\bar{\psi}_{\ell-m}^{(1,1)}(r,\theta)\right]
    +\bar{\eta}_{\ell m}\omega_{\ell m}^{(1)}\partial_\omega H_{0,\ell m}
    \left[{}_{2}\psi_{\ell m}^{(0,1)}(r,\theta)\right] \\
    =& \;\bar{\eta}_{\ell m}\bar{P}_{\ell-m}(\pi-\theta)
    \left[{}_{2}\psi_{\ell m}^{(0,1)}(r,\theta)\right]
    +\bar{Q}_{\ell m}(\pi-\theta)
    \left[{}_{2}\bar{\psi}_{\ell -m}^{(0,1)}(r,\theta)\right]\,,
\end{aligned}
\end{equation}    
\end{widetext}
where we have used Eqs.~\eqref{eq:omega1_symmetry}, \eqref{eq:parity_Teuk_func}, and that $\hat{\mathcal{P}}H_{0,\ell-m} = H_{0,\ell m}$ \cite{Li:2023ulk}. Notice that an additional factor of $(-1)^m$ is added to Eq.~\eqref{eq:master_eqn_scheme3_l-m} due to $\hat{\mathcal{P}}e^{im\phi}=(-1)^me^{im\phi}$. Although $e^{im\phi}$ does not show up in Eq.~\eqref{eq:master_eqn_scheme2_l-m}, it is necessary to keep track of this factor for self-consistency in the angular part of all the terms.

To solve for $\omega_{\ell m}^{(1)}$, one can construct an inner product, following \cite{Zimmerman:2014aha, Mark:2014aja, Hussain:2022ins}, that makes the Teukolsky operator in GR self-adjoint, i.e., 
\begin{equation} \label{eq:self_adjoint_product}
    \langle H_{0,\ell m}\varsigma(r,\theta)|\varphi(r,\theta)\rangle
    =\langle \varsigma(r,\theta)|H_{0,\ell m}\varphi(r,\theta)\rangle\,,
\end{equation}
where $\varsigma(r,\theta)$ and $\varphi(r,\theta)$ are some general functions in $(r,\theta)$ with the same asymptotic behaviors as the GR QNMs. As shown in \cite{Zimmerman:2014aha, Mark:2014aja}, the inner product in Eq.~\eqref{eq:self_adjoint_product} can be defined as an integral along certain contour $\mathscr{C}$, where
\begin{equation} \label{eq:inner_product_def}
    \langle\varsigma(r,\theta)|\varphi(r,\theta)\rangle
    =\int_{\mathscr{C}}\Delta^2(r)dr\int \sin\theta 
    \varsigma(r,\theta)\varphi(r,\theta) d\theta\,,
\end{equation}
and we will show below how to construct a contour $\mathscr{C}$ for a bumpy BH. Since $H_{0,\ell m}\left[{}_{2}\psi_{\ell m}^{(0,1)}(r,\theta)\right]=0$, if we take the inner product of ${}_{2}\psi_{\ell m}^{(0,1)}(r,\theta)$ with Eqs.~\eqref{eq:master_eqn_scheme2_lm} and \eqref{eq:master_eqn_scheme3_l-m} and use the property in Eq.~\eqref{eq:self_adjoint_product}, we can remove the first term on the left-hand side of Eqs.~\eqref{eq:master_eqn_scheme2_lm} and \eqref{eq:master_eqn_scheme3_l-m} such that this system of equations becomes a standard eigenvalue problem 
\begin{widetext}
\begin{equation} \label{eq:matrix_general}
    \frac{1}{\langle \partial_{\omega}H_{0,\ell m}\rangle}
    \begin{pmatrix}
        \langle P_{\ell m}\rangle
        & (-1)^{\ell}\langle Q_{\ell -m}\hat{\mathcal{C}}\hat{\mathcal{P}}\rangle \\
        (-1)^{\ell}\langle \bar{Q}_{\ell m}(\pi-\theta)
        \hat{\mathcal{C}}\hat{\mathcal{P}}\rangle
        & \langle \bar{P}_{\ell-m}(\pi-\theta)\rangle
    \end{pmatrix}
    \begin{pmatrix}
        1 \\ \bar{\eta}_{\ell m}
    \end{pmatrix}
    =\omega_{lm}^{(1)}\begin{pmatrix}
        1 \\ \bar{\eta}_{\ell m}
    \end{pmatrix}\,,
\end{equation}
\end{widetext}
where we have defined the shorthand notation
\begin{equation} \label{eq:inner_product_shorthand}
    \langle \mathcal{O}\rangle
    =\langle\psi_{\ell m}^{(0,1)}|\mathcal{O}
    \psi^{(0,1)}_{\ell m}\rangle\,.
\end{equation}
The matrix in Eq.~\eqref{eq:matrix_general} is the same as the one in Eq.~(68) of \cite{Li:2023ulk}, where one can directly map $P_{\ell m}$ and $Q_{\ell m}$ to the $(\ell,m)$ mode of $\mathcal{S}^{\mu\nu}\mathcal{O}_{\mu\nu}$ and $\mathcal{S}^{\mu\nu}\bar{\mathcal{O}}_{\mu\nu}$, respectively. The additional factor of $(-1)^{\ell}$ in the off-diagonal terms comes from that we choose to solve the pair of $\Psi_{0,\ell m}$ and $\Psi_{0,\ell -m}$ instead of $\Psi_{0,\ell m}$ and $\hat{\mathcal{P}}\Psi_{0,\ell m}$ in \cite{Li:2023ulk}, where $\hat{\mathcal{P}}\Psi_{0,\ell m}^{(0,1)}=(-1)^{\ell}\Psi_{0,\ell -m}$. Nonetheless, this choice and the resulting factor of $(-1)^{\ell}$ will not affect $\omega^{(1)}_{\ell m}$, as shown below. The solutions to Eq.~\eqref{eq:matrix_general} can be found in \cite{Li:2023ulk}, where the QNM frequencies are 
\begin{widetext}
\begin{equation} \label{eq:QNM_freq}
\begin{aligned}
    \omega_{\ell m}^{\pm(1)}
    =& \;\frac{\Big\langle P_{\ell m}
    +\bar{P}_{\ell-m}(\pi-\theta)\Big\rangle
    \pm\sqrt{\Big\langle P_{\ell m}
    -\bar{P}_{\ell-m}(\pi-\theta)\Big\rangle^2
    +4\Big\langle Q_{\ell-m}
    \hat{\mathcal{C}}\hat{\mathcal{P}}\Big\rangle
    \Big\langle \bar{Q}_{\ell m}(\pi-\theta)\hat{\mathcal{C}}
    \hat{\mathcal{P}}\Big\rangle}}
    {2\langle\partial_{\omega}H_{0,\ell m}\rangle}\,,
\end{aligned}
\end{equation}
and the coefficients $\eta_{\ell m}$ are
\begin{equation} \label{eq:eta_solns}
    \bar{\eta}_{\ell m}^{\pm}
    =(-1)^{\ell}\frac{\Big\langle\bar{P}_{\ell-m}(\pi-\theta)
    -P_{\ell m}\Big\rangle
    \pm\sqrt{\Big\langle P_{\ell m}
    -\bar{P}_{\ell-m}(\pi-\theta)\Big\rangle^2
    +4\Big\langle Q_{\ell-m}
    \hat{\mathcal{C}}\hat{\mathcal{P}}\Big\rangle
    \Big\langle \bar{Q}_{\ell m}(\pi-\theta)\hat{\mathcal{C}}
    \hat{\mathcal{P}}\Big\rangle}}
    {2\left\langle Q_{\ell -m}\hat{\mathcal{C}}\hat{\mathcal{P}}
    \right\rangle}\,.
\end{equation}
\end{widetext}
Notice, the factor of $(-1)^{\ell}$ in Eq.~\eqref{eq:eta_solns} disappears if one solve the pair of $\Psi_{0,\ell m}$ and $\hat{\mathcal{P}}\Psi_{0,\ell m}$ instead. As shown in \cite{Li:2023ulk}, in the special case that $\eta_{\ell m}^{\pm}=\pm 1$, the modified QNMs $\Psi_{0,\ell m}^{\eta(1,1)}$ are even- and odd-parity modes, respectively.

\subsection*{The contour for the bumpy BH}

To evaluate the inner product in Eq.~\eqref{eq:inner_product_def}, we must choose a contour $\mathscr{C}$ and impose correct boundary conditions so that the Teukolsky operator is self-adjoint. The boundary conditions of the QNMs require
\begin{align} \label{eq:GR_boundary_condition}
    & {}_{2}R_{\ell m}^{(0,1)}(r)\sim 
    r^{-5}e^{i\omega_{\ell m}r^{*}}\,,\;
    && r^{*}\to\infty\,, \nonumber\\
    & {}_{2}R_{\ell m}^{(0,1)}(r)\sim 
    \Delta^{-2}(r)e^{-i[\omega_{\ell m}-am/(2Mr_{+})]r^{*}}\,,\;
    && r^{*}\to-\infty\,,
\end{align}
Here, $r^{*}$ is the tortoise coordinate, where $r^{*}=\infty$ and $r^{*}=-\infty$ correspond to $r=\infty$ and $r=r_{+}$, respectively, with $r_{+}$ being the outer horizon of the Kerr spacetime. For Schwarzschild BHs, $r_{+}=2M$. Since $\text{Im}(\omega_{\ell m})<0$, ${}_{2}R_{\ell m}^{(0,1)}(r)$ diverges at both the horizon and the real infinity. However, a finite inner product can still be constructed by considering a contour $\mathscr{C}$ in the complex plane and analytically continuing the radial Teukolsky functions. Consider a contour with endpoints at $r=a$ and $r=b$ in the complex plane. We need to ensure that the Teukolsky operator is self-adjoint. Evaluating Eq.~\eqref{eq:self_adjoint_product} with Eq.~\eqref{eq:inner_product_def}, one can first carry out the angular integral by projecting both $\varsigma(r,\theta)$ and $\varsigma(r,\theta)$ to spin-weighted spheroidal harmonics, i.e.,
\begin{equation}
    \varsigma(r,\theta)
    =\sum_{\ell m}\varsigma_{\ell m}(r){}_{2}
    S_{\ell m}(\theta)\,,
\end{equation}
and similarly for $\varphi(r,\theta)$. Then one can use the radial Teukolsky operator given by Eq.~\eqref{eq:radial_Teukolsky_eqn} and perform the radial integral in Eq.~\eqref{eq:inner_product_def}. After integrating by parts, one is left with
\begin{equation} \label{eq:boundary_terms}
    \Delta^{3}(r)[\partial_r\varsigma_{\ell m}(r)\varphi_{\ell m}(r)
    -\partial_r\varphi_{\ell m}(r)\varsigma_{\ell m}(r)]\bigg|^{b}_{a}\,.
\end{equation}
If one follows \cite{Zimmerman:2014aha, Mark:2014aja, Hussain:2022ins} to choose the contour $\mathscr{C}_1$ in Fig.~\ref{fig:NewContour}, which surrounds the imaginary axis at $r_{+}$, the boundary terms in Eq.~\eqref{eq:boundary_terms} vanish, so Eq.~\eqref{eq:self_adjoint_product} is satisfied. It is because we have imposed that $\varsigma_{\ell m}(r)$ and $\varphi_{\ell m}(r)$ satisfy the same boundary conditions as ${}_{2}R_{\ell m}^{(0,1)}(r)$, while ${}_{2}R^{(0,1)}_{\ell m}(r)\to 0$ as $r\to r_{+}\pm\epsilon+i\infty$, where $\epsilon\ll M$ and $\epsilon$ is real.

\begin{figure}[t]
    \includegraphics[width=0.5\textwidth]{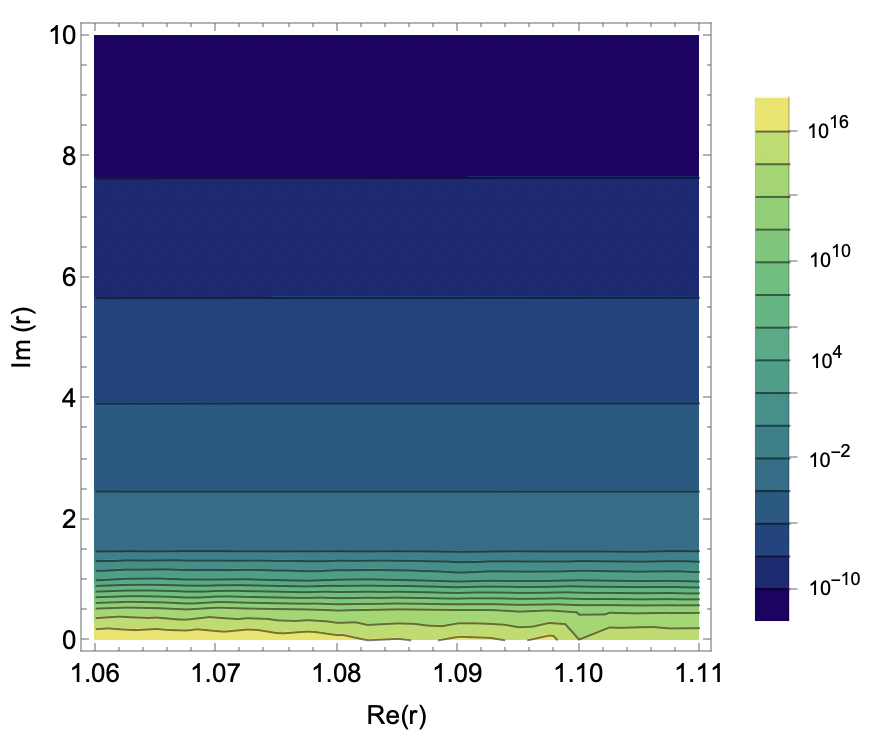}
    \caption{The absolute value of $\langle P_{\ell m}\rangle$ after the angular integral in Eq.~\eqref{eq:inner_product_def} is shown for the $(\ell,m,n)=(2,1,0)$ mode and the $\ell_W=2$ Weyl multipole. We use the convention that $M=1/2$ in \cite{Leaver:1985ax} in this plot.}
    \label{fig:EffPotential}
\end{figure}

Although the contour $\mathscr{C}_1$ works for many cases, it does not really apply to the modified Teukolsky equations for the bumpy BHs here. The main reason is that certain NP quantities at $\mathcal{O}(\zeta^1,\epsilon^0)$ diverge when $\text{Re}(r)\leq2M$ due to the distance function $d(r,\theta)$ defined in Eq.~\eqref{eq:DistanceFunc}. In this case, the radial integral along the half of the contour $\mathscr{C}_1$ inside the horizon diverges. To avoid this issue, we consider an alternative contour $\mathscr{C}_2$, which starts from $r=2M+\epsilon$, where $\epsilon< M$ and $\epsilon$ is real, and ends at $r=2M+\epsilon+i\infty$, as depicted in Fig.~\ref{fig:NewContour}. To ensure that all the terms in Eq.~\eqref{eq:matrix_general} are well-behaved along $\mathscr{C}_2$, we perform the angular integral in Eq.~\eqref{eq:inner_product_def} of the source terms and inspect their absolute value for complex-valued $r$ around $\mathscr{C}_2$. As an example, we show in Fig.~\ref{fig:EffPotential} that $\langle P_{\ell m}\rangle$ is non-singular along $\mathscr{C}_2$. 

According to Eq.~\eqref{eq:boundary_terms}, to ensure that the Teukolsky operator is still self-adjoint, or at least that we can  remove the first term on the left-hand side of Eqs.~\eqref{eq:master_eqn_scheme2_lm} and \eqref{eq:master_eqn_scheme3_l-m}, we need to impose
\begin{equation}\label{eq:BoundaryCond}
   \frac{{}_{2}R_{\ell m}^{(1,1)}(r)}
    {\partial_r\left({}_{2}R_{\ell m}^{(1,1)}(r)\right)}
    =\frac{{}_{2}R_{\ell m}^{(0,1)}(r)}
    {\partial_r\left({}_{2}R_{\ell m}^{(0,1)}(r)\right)}
\end{equation}
at both $r=2M+\epsilon$ and $r=2M+\epsilon+i\infty$. The condition at $r=2M+\epsilon+i\infty$ is easily satisfied since the source terms in Eqs.~\eqref{eq:master_eqn_scheme2_lm} and \eqref{eq:master_eqn_scheme3_l-m} vanish as $r\to2M+\epsilon+i\infty$. However, the condition at $r=2M+\epsilon$ is not naturally satisfied since the source terms do not vanish near the horizon, resulting in nonzero ${}_{2}R_{\ell m}^{(0,1)}(r)$ and ${}_{2}R_{\ell m}^{(1,1)}(r)$.

\begin{figure*}[t]
    \centering
    \includegraphics[width=0.7\textwidth]{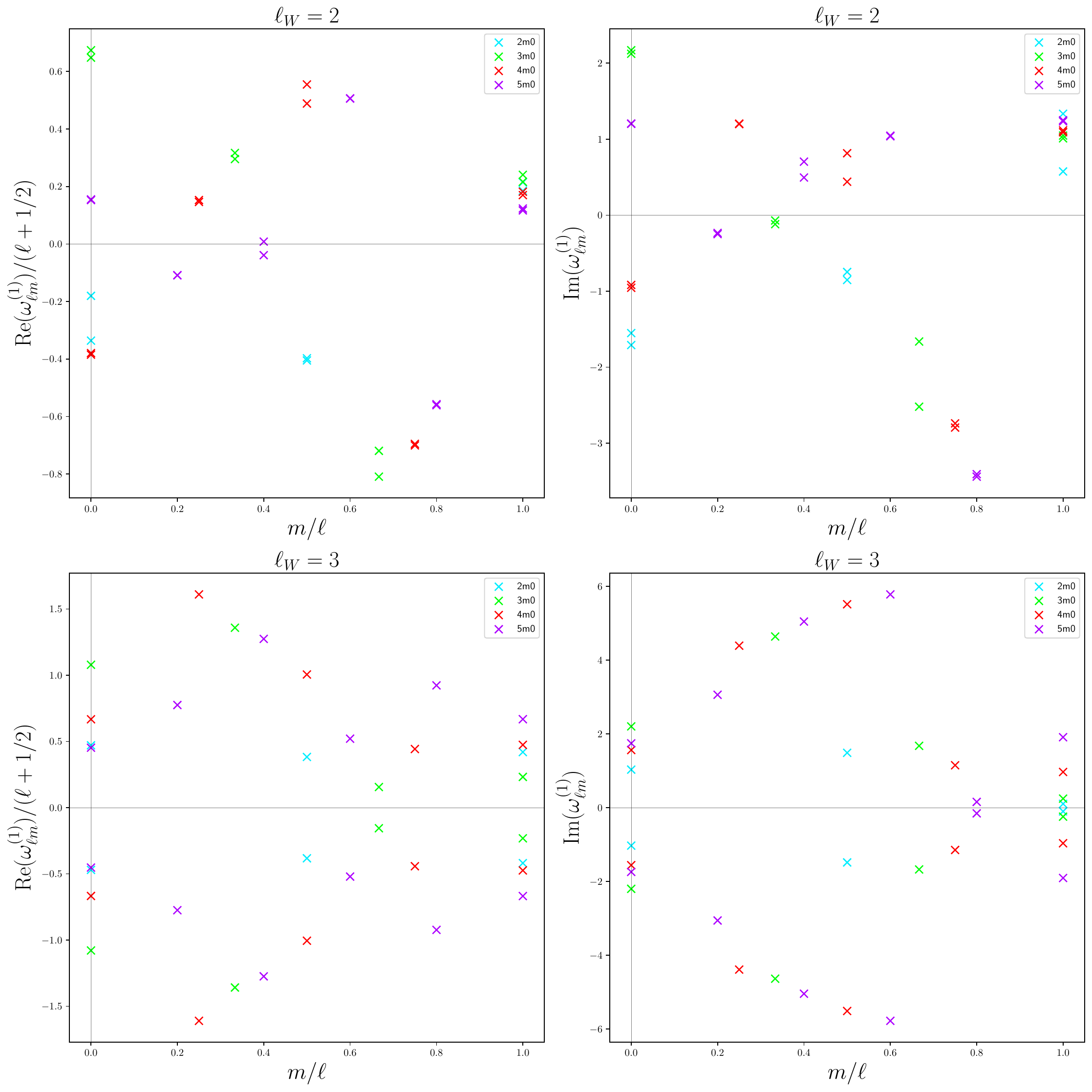}
    \caption{The real (left column) and imaginary (right column) parts of the QNM frequency shifts $\omega^{\pm(1)}_{\ell m}$ generated by the $\ell_W=2$ bump (top row) and $\ell_W=3$ bump (bottom row) are presented. For each frequency, the fundamental mode $n=0$ is shown. For simplicity, we use the same marker for $\omega^{+(1)}_{\ell m}$ and $\omega^{-(1)}_{\ell m}$ for each $(\ell,m)$ mode.}
    \label{fig:modes}
\end{figure*}

One resolution is to consider the ``membrane paradigm" in \cite{Thorne:1986iy, Price:1986yy, Damour:1978cg, Znajek_1978}. This formalism elucidates the physical nature of the BH horizon by modeling it as a fictitious fluid membrane. The dynamics of the membrane are parametrized by a dissipative stress-energy tensor that sets the fluid's velocity, density, pressure, shear viscosity, and bulk viscosity. For example, it was found for a Schwarzschild BH that it has bulk viscosity $\zeta=-1/16\pi$ and shear viscosity $\eta=1/16\pi$. Altering these transport coefficients would alter the near horizon geometry and, therefore, modify the boundary conditions of the resulting QNMs \cite{Maggio:2020jml, Chakraborty:2022zlq}. For example, one can have some nonzero reflectivity at the membrane, generating GW echos \cite{Chen:2020htz}. One may then deliberately pick some fluid stress tensor such that the spacetime is a bumpy BH for $r>2M+\epsilon$, while the spacetime is still a Schwarzschild spacetime for $2M\leq r\leq 2M+\epsilon$. In this case, we can impose the condition in Eq.~\eqref{eq:BoundaryCond} at $r=2M+\epsilon$ since ${}_{2}R_{\ell m}^{(1,1)}$ satisfies the same boundary condition of a GR QNM [i.e., Eq.~\eqref{eq:GR_boundary_condition}]. In this work, we do not explicitly provide a stress tensor giving rise to Eq.~\eqref{eq:BoundaryCond} but assume its possible existence. We will strictly derive the relation between the fluid stress tensor and the boundary condition of the QNMs at the membrane in our future work. In general, different fluids can result in different boundary conditions other than Eq.~\eqref{eq:GR_boundary_condition}, for example, a nonzero reflectivity at the membrane. This nonzero reflectivity can make additional modifications to the QNM spectrum \cite{Chen:2020htz}. Since the boundary condition in Eq.~\eqref{eq:BoundaryCond} is the most natural to use, we will stick to it for the rest of this work.

\section{Results}

\label{sec:results}

\begin{figure*}[t]
\includegraphics[width=0.7\textwidth]{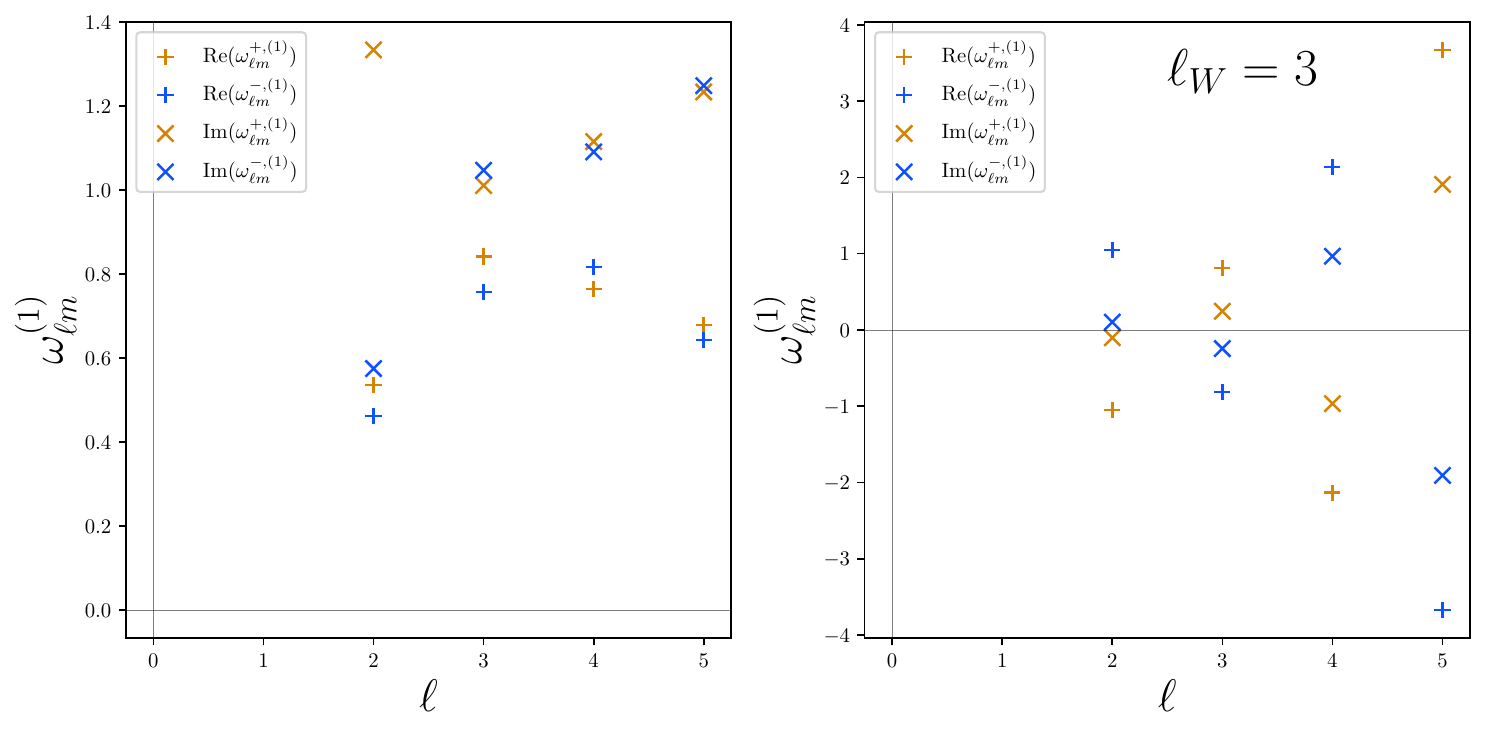}
    \caption{The real and imaginary parts of the QNM frequency shifts $\omega^{\pm(1)}_{\ell m}$ for $n=0$, $\ell=m$ up to $\ell=5$ for both the $\ell_W=2$ (left panel) and $\ell_W=3$ (right panel) Weyl multipole corrections.}
    \label{fig:ReAndImRegression}
\end{figure*}

Using the contour $\mathscr{C}_2$ in Sec.~\ref{sec:EVP} and Fig.~\ref{fig:NewContour}, we can now compute the QNM frequency shifts generated by a bumpy BH. The contour $\mathscr{C}_2$ was chosen by setting $\epsilon=0.2M$. All the $\mathcal{O}(\zeta^0,\epsilon^1)$ quantities were computed using Leaver's method \cite{Leaver:1985ax}. The results of the QNM frequency shifts are shown in Fig.~\ref{fig:modes}, where both the real and imaginary parts of the QNM frequency shifts $\omega^{\pm(1)}_{\ell m}$ are plotted. Specifically, for both the Weyl bumps $\ell_W=2$ and $\ell_W=3$, we plot all the modes $(\ell,m,n)$ with $\ell=2,3$, $m\geq0$, and $n=0$. The frequencies of the modes with $m<0$ can be found using the relation in Eq.~\eqref{eq:omega1_symmetry}. The QNM frequency shifts $\omega^{\pm(1)}_{\ell m}$ of any additional overtones we have calculated are listed in the tables in Appendix~\ref{appendix:qnm_tables}.

In Fig.~\ref{fig:modes}, one important feature is that for each $(\ell,m)$ mode, the frequency shift is degenerate for the Weyl bump $\ell_W=3$, i.e., $\omega^{+(1)}_{\ell m}=-\omega^{-(1)}_{\ell m}$. In contrast, there are two independent shifts for the Weyl bump $\ell_W=2$. This is a natural consequence of bumps' parity. Notice that the metric correction $h^{(1,0)}_{\mu\nu}$ in \eqref{eq:LinearPerturbMetric} for each set of potentials $(\psi^{\ell_W}_1,\gamma^{\ell_W}_1)$ obeys
\begin{equation}
    \hat{\mathcal{P}}h^{(1,0)}_{\mu\nu}
    =(-1)^{\ell_W}h^{(1,0)}_{\mu\nu}\,,
\end{equation}
as one can explicitly check using Eqs.~\eqref{eq:psi_l2}, \eqref{eq:gamma_l2}, and \eqref{eq:bumps_l3} for $\ell_W=2,3$. 
We can then derive the $\hat{\mathcal{P}}$ transformation of all the NP quantities at $\mathcal{O}(\zeta^1,\epsilon^0)$ following \cite{Li:2023ulk}. In the end, we get that $\hat{\mathcal{P}} H^{(1,0)}_{0}=(-1)^{\ell_W}H^{(1,0)}_{0}$, which implies
\begin{align}
    & \langle P_{\ell m}\rangle
    =(-1)^{\ell_W}\langle\bar{P}_{\ell-m}(\pi-\theta)\rangle\,, 
    \nonumber\\
    & \langle Q_{\ell-m}\hat{\mathcal{C}}\hat{\mathcal{P}}\rangle
    =(-1)^{\ell_W}\langle\bar{Q}_{\ell m}(\pi-\theta)
    \hat{\mathcal{C}}\hat{\mathcal{P}}\rangle\,.
\end{align}
Using Eq.~\eqref{eq:QNM_freq}, we can derive two salient relations governing the QNM frequency shifts $\omega_{\ell m}^{\pm (1)}$. First, for the odd-parity Weyl multipoles, there is an additional symmetry of isospectrality breaking, where $\omega^{-(1)}_{\ell m}=-\omega^{+(1)}_{\ell m}$. Even though we use the notation $+$ and $-$ to label the frequency shifts, the resulting wavefunctions do not have definite parity since $\eta_{\ell m}^{\pm}\neq\pm1$ \cite{Li:2023ulk}. Second, for the even-parity Weyl multipoles, Eq.~\eqref{eq:eta_solns} implies that the resulting wavefunction has definite parity, i.e., $\eta_{\ell m}^{\pm}=\pm1$ \cite{Li:2023ulk}. These features were also observed in \cite{Cano:2021myl, Cano:2023tmv, Cano:2023jbk} when studying the QNM frequency shifts of higher-derivative gravity for both the parity-preserving and parity-violating corrections to the Einstein-Hilbert action. In \cite{Li:2023ulk} and this work, we more directly show the origin of these features. Nonetheless, isospectrality is broken for all the modes we have calculated regardless of the parity of the bumps.

Furthermore, it was first shown in \cite{Ferrari:1984zz} that for a Schwarzschild BH, the real part of the QNM frequencies is related to the orbital frequencies of null geodesics near the light ring, while the imaginary part encodes the Lyapunov exponent of the orbit. Specifically, when $m=\ell$, the QNM frequencies $\omega_{\ell mn}$ have the following eikonal approximation,
\begin{equation} \label{eq:EikonalApprox}
    \omega_{\ell mn}
    \approx(\ell+1/2)\Omega-i\gamma_L(n+1/2)\,,
\end{equation}
where $\Omega$ is the Keplerian frequency of a circular null geodesic, and $\gamma_L$ is the Lyapunov exponent. For an axisymmetric spacetime without spherical symmetry, one would naively expect that the real part of $\omega_{\ell m n}$ for fixed low values of $n$ will depend linearly on $\ell$ for modes with the same values of $m/\ell$, which corresponds to the inclination angle of the orbit the mode is associated with. The imaginary part of $\omega_{\ell mn}$ for low values of $n$ are expected to stay roughly constant as $\ell$ increases while fixing $m/\ell$ and $n$, since these values are related to the Lyapunov exponents of the orbits. However, as shown in Fig.~\ref{fig:ReAndImRegression}, the real part $\operatorname{Re}(\omega^{(1)}_{\ell m})$ does not depend on $\ell$ linearly, which is inconsistent with the prediction in Eq.~\eqref{eq:EikonalApprox}. Moreover, the imaginary part $\operatorname{Im}(\omega^{(1)}_{\ell m})$ is not constant for the mode $n=0$. These inconsistencies suggest that the relationship predicted by the eikonal approximation in Eq.~\eqref{eq:EikonalApprox} may need further exploration for these bumpy BHs and BHs in bGR theories in general.

In general, one may not be able to use observational data to examine the QNM frequency shifts for each Weyl bump independently. In this case, we need to sum the contributions to $\omega^{(1)}_{\ell m}$ from bumps with different $\ell_W$. As we previously discussed, the bumps with odd $\ell_W$ and even $\ell_W$ have different parity, so one needs to use the recombination rule found in \cite{Cano:2021myl} in general, i.e., 
\begin{widetext}
\begin{equation}
     \omega^{\pm(1)}_{\text{total},\ell m}
     =\frac{\omega^{+,(1)}_{\even,\ell m}
     +\omega^{-(1)}_{\even,\ell m}}{2}
     \pm\sqrt{\left(\frac{\omega^{+(1)}_{\even,\ell m}
     -\omega^{-(1)}_{\even,\ell m}}{2}\right)^2
     +\left(\omega^{+(1)}_{\odd,\ell m}\right)^2}\,,
\end{equation}
\end{widetext}
where $\omega^{\pm(1)}_{\even,\ell m}$ and $\omega^{\pm(1)}_{\odd,\ell m}$ are the QNM frequency shifts generated by the even-parity (i.e., even $\ell_W$) and odd-parity (i.e., odd $\ell_W$) Weyl multipoles, respectively.

\section{Conclusion and Outlook}
\label{sec:conclusion}

In this work, we used the modified Teukolsky formalism in \cite{Li:2022pcy, Li:2023ulk} to compute the QNM frequency shifts for a non-rotating BH with axisymmetric deviations parametrized by Weyl multipoles. Since these bumpy BHs are Ricci flat in the non-rotating case, and we did not consider any corrections to the Einstein-Hilbert action, the only term in the modified Teukolsky equation that contributes is $\mathcal{S}_{\geo}^{(1,1)}$, which only depends on the modifications to the background geometry [i.e., terms at $\mathcal{O}(\zeta^1,\epsilon^0)$] and the QNMs in GR [i.e., terms at $\mathcal{O}(\zeta^0,\epsilon^1)$]. The terms at $\mathcal{O}(\zeta^1,\epsilon^0)$ were directly evaluated using the bumpy BH metric found in \cite{Vigeland:2009pr}. To calculate the $\mathcal{O}(\zeta^0,\epsilon^1)$ quantities, we implemented the CCK-Ori metric reconstruction procedure in the IRG. After obtaining the modified Teukolsky equations, we noticed that the source terms mix the $(\ell,m)$ and $(\ell,-m)$ modes, which is one main cause for isospectrality breaking \cite{Li:2023ulk}. Following \cite{Hussain:2022ins, Li:2023ulk}, we solved the $(\ell,m)$ and $(\ell,-m)$ modes jointly and used the EVP method in \cite{Zimmerman:2014aha, Mark:2014aja, Hussain:2022ins} to compute the QNM frequencies.   

We obtained the QNM frequency shifts for the modes $\ell=2,3$ up to the second overtone for both the bumps with multipole $\ell_W=2$ and $\ell_W=3$. Some qualitative features were found. Our results showed that isospectrality is broken for both the $\ell_W=2$ and $\ell_W=3$ bumps. Specifically, we noticed that the isospectrality breaking structure is related to the parity of the bumps. For odd-parity bumps, we found that the two frequency shifts, due to isospectrality breaking, are opposite to each other. These features are consistent with the ones discovered by \cite{Cano:2021myl, Cano:2023tmv, Cano:2023jbk} in higher-derivative gravity. Furthermore, we notice that the eikonal approximation in GR \cite{Ferrari:1984zz, Cardoso:2008bp, Yang:2012he} is no longer valid here, so further investigation is needed in these bumpy BH spacetimes. We suspect this breakdown can be related to the chaotic nature of photon orbits near bumpy BHs, as found by Refs.~\cite{Glampedakis:2018blj,Rojas:2022}.

Since we expect to observe the GW ringdown with a much higher signal-to-noise ratio during the fourth LVK observing run and in the future with third-generation detectors, such as Einstein Telescope \cite{Punturo:2010zz} and Cosmic Explorer \cite{Dwyer:2014fpa}, it is critical to accurately model the GWs emitted during the ringdown phase not only in GR but also bGR theories. Our work has made a crucial attempt in this direction by developing a framework to study the ringdown of a BH spacetime with parametrized deviations, which is valid for rotating BHs in general. Through our work, one can directly connect the multipole structure of a BH spacetime to its QNM spectra.

Our efforts can be extended in several directions. In \cite{Vigeland:2009pr}, Hughes and Vigeland also derived a spinning bumpy BH spacetime. Since the bumps in the non-rotating case are already axisymmetric, the procedures in this work should naturally extend to spinning bumpy BHs. One subtlety is that the bumpy spinning BH in \cite{Vigeland:2009pr} is not Ricci flat, unlike the non-spinning case. This introduces extra source terms driven by the Ricci tensor and not present in our current implementation. Even though these source terms decay as $r^{-(\ell_W+1)}$ for the Weyl multipole $\ell_W$, we still need to investigate whether they significantly contribute to the QNM frequency shifts. Fortunately, the modified Teukolsky formalism can still deal with these extra source terms, as demonstrated in \cite{Cano:2023tmv, Cano:2023jbk, Wagle:2023fwl}. Another theory-agnostic approach could instead use the GH multipole moments, which can parametrize Ricci-flat solutions to the Einstein equations even for rotating BHs. A potential drawback to this strategy is that one cannot uniquely describe a non-vacuum spacetime with multipoles \cite{Bonga:2021ouq}, limiting the ability to model more general BH environments \cite {Speri:2022upm}. Nonetheless, multipoles can still be a valuable probe of possible bGR corrections and a clean way to formulate no-hair tests of GR.

Since isospectrality breaking is likely a common feature in bGR theories \cite{Li:2023ulk}, understanding its physical origins and signatures on GW observables is another important direction. In this work, we have made a direct connection from the parity of the bumps to the isospectrality breaking structure of the QNMs. Since the same relations also appeared in higher-derivative gravity \cite{Cano:2021myl,Cano:2023tmv, Cano:2023jbk} and other parametrized ringdown studies of non-rotating BHs \cite{McManus:2019ulj}, it will be worth to derive these relations from the parity properties of the action directly following \cite{Li:2023ulk}. Furthermore, we should also investigate how these generic isospectrality-breaking features of QNMs impact GW waveforms. We should study whether we can extract any characteristic features associated with isospectrality breaking from the waveforms. These features are potentially useful for testing the class of parity-preserving or parity-violating theories in a generic way.


\acknowledgements   

We thank Scott Hughes and Andrew Laeuger for useful discussions.  C.W., D.L., and Y.C.'s research is supported by the Simons Foundation (Award No. 568762), the Brinson Foundation, and the National Science Foundation (via Grants No. PHY-2011961 and No. PHY-2011968).

\appendix
\section{$\mathcal{O}(\zeta^1, \epsilon^0)$ quantities}
\label{appendix:order_10_quantities}
In this section, we list the Weyl scalars and spin coefficients at $\mathcal{O}(\zeta^1,\epsilon^0)$. The Weyl scalars at $\mathcal{O}(\zeta^1,\epsilon^0)$ are given by
\begin{widetext}
\begin{align}
    \Psi^{(1,0)}_0
    =& \;-\frac{\partial_{\theta}^2\gamma_1
    +(r-2M)\left(2\partial_r+r\partial_r^2\right)\gamma_1
    -\cot{\theta}\partial_{\theta}\gamma_1
    -2\partial_{\theta}^2\psi_1
    +2\cot{\theta}\partial_{\theta}\psi_1}{2r(r-2M)}\,, 
    \label{eq:psi0_bumpy} \\
    \Psi^{(1,0)}_1
    =& \;\frac{2r(r-2M)\left(\cot{\theta}\partial_r\gamma_1
    -2\partial_r\partial_{\theta}\psi_1\right)
    +2(r+M)\partial_{\theta}\gamma_1
    +4(r-5M)\partial_{\theta}\psi_1}{4\sqrt{2}r^2(r-2M)}\,, \\
    \Psi^{(1,0)}_2
    =& \;\frac{r(r-2M)\partial^2_r
    \left(8\psi_1-5\gamma_1\right)-2(2M+r)\partial_r\gamma_1
    +4(M+r)\partial_r\psi_1}{12r^2}\\ \nonumber
    & \;+\frac{r\left[\cot{\theta}\left(2\partial_{\theta}\psi_1
    -3\partial_{\theta}\gamma_1\right)-5\partial^2_{\theta}\gamma_1
    +2\partial^2_{\theta}\psi_1\right]+24M\gamma_1-24M\psi_1}{12r^3}\,, 
    \label{eq:psi2_bumpy} \\
    \Psi^{(1,0)}_3
    =& \;\frac{-2r(r-2M)\partial_r\left(\cot{\theta}\gamma_1
    -2\partial_{\theta}\psi_1\right)
    -2(r+M)\partial_{\theta}\gamma_1
    +4(5M-r)\partial_{\theta}\psi_1}{8\sqrt{2}r^3}\,, \\
    \Psi_4^{(1,0)}
    =& \;\frac{(r-2M)\left[\cot{\theta}\partial_{\theta}
    \left(\gamma_1-2\psi_1\right)
    -\partial_\theta^2\gamma_1-(r-2M)
    \left(2\partial_r\gamma_1+r\partial_r^2\gamma_1\right)
    +2\partial_{\theta}^2\psi_1\right]}{8r^3}\,.
\end{align}
\end{widetext}
The spin coefficients at $\mathcal{O}(\zeta^1,\epsilon^0)$ are given by
\begin{subequations} \label{eq:spin_coeffs_bumpy}
\begin{align}
    & \kappa^{(1,0)}
    =\frac{\partial_{\theta}\gamma_1-2\partial_{\theta}\psi_1}
    {\sqrt{2}(r-2M)}\,, \\
    & \pi^{(1,0)}
    =\frac{\partial_{\theta}\gamma_1}{2\sqrt{2}r}\,,\\
    & \epsilon^{(1,0)}
    =\frac{1}{2}\partial_r\psi_1\,, \\
    & \rho^{(1,0)}
    =\frac{-r\partial_r\gamma_1+2\gamma_1
    +2r\partial_r\psi_1-2\psi_1}{2r}\,, \\
    & \lambda^{(1,0)}
    =-\frac{(r-2M)\partial_r\gamma_1}{4r}\,, \\
    & \alpha^{(1,0)}
    =\frac{\cot{\theta}(\gamma_1-\psi_1)+\psi_1}{2\sqrt{2}r}\,, \\
    & \sigma^{(1,0)}
    =-\frac{1}{2}\partial_r\gamma_1\,, \\
    & \mu^{(1,0)}
    =-\frac{(r-2M)\left[r\left(\partial_r\gamma_1
    -2\partial_r\psi_1\right)-2\gamma_1+2\psi_1\right]}{4r^2}\,, \\
    & \beta^{(1,0)}
    = -\frac{\cot{\theta}\left(\gamma_1-\psi_1\right)
    +\partial_\theta\psi_1}{2\sqrt{2}r}\,, \\
    & \nu^{(1,0)} 
    = -\frac{(r-2M)\left(\partial_{\theta}\gamma_1
    -2\partial_{\theta}\psi_1\right)}{4\sqrt{2}r^2}\,, \\
    & \gamma^{(1,0)}
    =\frac{-2M\gamma+r(r-2M)\partial_r\psi_1+2M\psi_1}{4r^2}\,, \\
    & \tau^{(1,0)}
    =-\frac{\partial_\theta\gamma_1}{2\sqrt{2}r}\,.
\end{align}    
\end{subequations}

\section{QNM Frequency Shifts}
\label{appendix:qnm_tables}

In this appendix, we explicitly tabulate the QNM frequency shifts $\omega^{\pm(1)}_{\ell m}$ generated by the Weyl multipoles $\ell_W=2,3$. All the results in this appendix were calculated using the modified contour $\mathscr{C}_2$ in Fig.~\ref{fig:NewContour}. For the even-parity bump $\ell_W=2$, the top row of each cell refers to $\omega^{+(1)}_{\ell m}$ with $\eta_{\ell m}=(-1)^{\ell}$, while the bottom row of each cell refers to $\omega^{-(1)}_{\ell m}$ with $\eta_{\ell m}=(-1)^{\ell+1}$. For the odd-parity bump $\ell_W=3$, since $\omega^{-(1)}_{\ell m}=-\omega^{+(1)}_{\ell m}$ as discussed in Sec.~\ref{sec:results}, we only list $\omega^{+(1)}_{\ell m}$. 
\begin{widetext}
\begin{center}
\begin{table}[h!]
    \centering
    \caption{$\ell=2\,,\ell_W=2$}
    \begin{tabular}{||c|c|c|c||}
    \hline
    &$m=2$ & $m=1$ & $m=0$ \\
    \hline\hline
    \multirow{2}{4em}{$n=0$} 
 &$0.5356+1.334i$ & $-0.9934-0.7472i$ & $-0.8402-1.552i$ \\
&$0.4624+0.5751i$ & $-1.013-0.8530i$ & $-0.4508-1.709i$ \\
    \hline
    \multirow{2}{4em}{$n=1$} 
&$-0.2501+1.209i$ & $-0.6474-1.133i$ & $-0.08632-1.464i$ \\
&$0.5693+1.013i$ & $-0.5678-1.158i$ & $-0.04923-1.735i$ \\
    \hline
    \multirow{2}{4em}{$n=2$} 
& $-0.1020+1.114i$ & $-0.3401-1.208i$ & $0.03257-1.560i$ \\
&$0.1645+1.208i$ & $-0.3330-1.162i$ & $0.08164-1.433i$ \\
\hline
    \end{tabular}
\end{table}

\begin{table}[h!]
    \centering
    \caption{$\ell=2\,,\ell_W=3$}
    \begin{tabular}{||c|c|c|c||}
    \hline
    & $m=2$ & $m=1$ & $m=0$ \\
    \hline \hline
    $n=0$ &$-1.048-0.1024i$ & $0.9566+1.485i$ & $1.176+1.030i$ \\
    $n=1$ & $0.05390+0.6246i$ & $-0.3629-1.547i$ & $-0.6855-1.265i$ \\
    $n=2$ & $0.08806-0.3659i$ & $-0.1525-1.440i$ & $-0.3984-1.171i$ \\
    \hline
    \end{tabular}
\end{table}

\begin{table}[h!]
    \centering
    \caption{$\ell=3\,,\ell_W=2$}
    \begin{tabular}{||c|c|c|c|c||}
    \hline
    & $m=3$ & $m=2$ & $m=1$ & $m=0$ \\
    \hline \hline
    \multirow{2}{4em}{$n=0$} 
    & $0.8419+1.011i$ & $-2.832-2.520i$ & $1.108-0.1184i$ & $2.358+2.124i$ \\
    & $0.7574+1.047i$ & $-2.518-1.663i$ & $1.034-0.06974i$ & $2.268+2.170i$ \\
    \hline 
    \multirow{2}{4em}{$n=1$} 
    & $0.4823+1.185i$ & $-1.862-3.176i$ & $1.124+0.5143i$ & $1.638+2.907i$ \\
    &$0.4446+1.118i$ & $-2.244-2.753i$ & $1.093+0.4545i$ & $1.602+2.840i$ \\
    \hline
    \multirow{2}{4em}{$n=2$} 
    & $0.3154+1.023i$ & $-1.323-2.965i$ & $0.8396+0.6965i$ & $1.099+2.720i$ \\
    &$0.2872+1.043i$ & $-1.562-2.898i$ & $0.8119+0.6996i$ & $1.072+2.724i$ \\
    \hline
    \end{tabular}
\end{table}

\begin{table}[h!]
    \centering
    \caption{$\ell=3\,,\ell_W=3$}
    \begin{tabular}{||c|c|c|c|c||}
    \hline
    &$m=3$ &$m=2$ & $m=1$ & $m=0$ \\
    \hline\hline
    $n=0$ & $0.8113+0.2449i$ & $-0.5446-1.675i$ & $-4.754-4.637i$ & $3.775+2.201i$\\
    $n= 1$ & $0.7898+0.8195i$ & $-0.0214-1.425i$ & $3.442+6.196i$ & $3.341+3.550i$\\
 $n=2$  &$0.5539+1.071i$ & $-0.05605+0.9151i$ & $2.324+6.015i$ & $2.443+3.666i$\\
    \hline
    \end{tabular}
\end{table}
\end{center}
\end{widetext}

\bibliographystyle{apsrev4-1}
\bibliography{reference}

\end{document}